\documentclass{mn2e}
\usepackage{graphicx} 
\usepackage{natbib}
\usepackage{placeins}
\usepackage{amsmath}
\usepackage{caption}
\usepackage{longtable}
\usepackage{pdflscape}
\usepackage{comment}
\usepackage{hyperref}

\hypersetup{colorlinks=false}

\makeatletter

\newcommand{\Rmnum}[1]{\expandafter\@slowromancap\romannumeral #1@}
\newcommand{\etal}{{\it et al.}}
\newcommand{\hi}{{\sc H\,i}}
\newcommand{\hii}{{\sc H\,ii}}
\makeatother

\begin{document}
\title{The Herschel Fornax Cluster Survey II: FIR properties of optically-selected Fornax cluster galaxies}
\author[Fuller \etal]
{C. Fuller$^{1}$\thanks{E-mail: \texttt{chris.fuller@astro.cf.ac.uk}},
J. I. Davies$^{1}$,
R. Auld$^{1}$,
M. W. L. Smith$^{1}$, 
M. Baes$^{2}$, 
S. Bianchi$^{3}$,
M. Bocchio$^{4}$,\newauthor
A. Boselli$^{5}$,  
M. Clemens$^{6}$,
T. A. Davis$^{7}$,
I. De Looze$^{2}$,
S. di Serego Alighieri$^{3}$,
M. Grossi$^{6}$,\newauthor
T. M. Hughes$^{2}$,
S. Viaene$^{2}$, and
P. Serra$^{8}$ \\ \\
$^{1}$School of Physics and Astronomy, Cardiff University, The Parade, Cardiff, CF24 3AA, UK. \\
$^{2}$Sterrenkundig Observatorium, Universiteit Gent, Krijgslaan 281 S9, B-9000 Gent,Belgium. \\
$^{3}$INAF-Osservatorio Astrofisico di Arcetri, Largo Enrico Fermi 5, 50125 Firenze, Italy. \\
$^{4}$Laboratoire AIM, CEA/DSM- CNRS - Universit\'e Paris Diderot, Irfu/Service, Paris, France. \\
$^{5}$Laboratoire d'Astrophysique de Marseille, UMR 6110 CNRS, 38 rue F. Joliot-Curie, F-13388 Marseille, France. \\
$^{6}$INAF-Osservatorio Astronomico di Padova, Vicolo dell'Osservatorio 5, 35122 Padova, Italy. \\
$^{7}$European Southern Observatory, Karl-Schwarzschild Str. 2, 85748 Garching bei Muenchen, Germany.  \\
$^{6}$CAAUL, Observat\'orio Astron\'omico de Lisboa, Universidade de Lisboa, Tapada da Ajuda, 1349-018, Lisboa, Portugal. \\
$^{8}$CSIRO Astronomy and Space Science, Australia Telescope National Facility, PO Box 76, Epping, NSW 1710, Australia.
 }

\date{\today}
\maketitle

\begin{abstract}
The $Herschel$ Fornax Cluster Survey (HeFoCS) is a deep, far-infrared (FIR) survey of the Fornax cluster. The survey is in 5 $Herschel$ bands (100 - 500 \micron) and covers an area of 16 deg$^2$ centred on NGC\,1399. This paper presents photometry, detection rates, dust masses and temperatures using an optically selected sample from the Fornax Cluster Catalogue (FCC). Our results are compared with those previously obtained using data from the $Herschel$ Virgo Cluster Survey (HeViCS). In Fornax, we detect 30 of the 237 (13\,\%) optically selected galaxies in at least one $Herschel$ band. The global detection rates are significantly lower than Virgo, reflecting the morphological make up of each cluster - Fornax has a lower fraction of late-type galaxies. For galaxies detected in at least 3 bands we fit a modified blackbody with a $\beta = 2$ emissivity. Detected early-type galaxies (E\,/\,S0) have a mean dust mass, temperature, and dust-to-stars ratio of $\log_{10}(<M_{dust}>/\mathrm{M_{\odot}}) = 5.82 \pm 0.20$, $<T_{dust}> = 20.82 \pm 1.77$\,K, and $\log_{10}(M_{dust}/M_{stars}) = -3.87 \pm 0.28$, respectively. Late-type galaxies (Sa to Sd) have a mean dust mass, temperature, and dust-to-stars ratio of $\log_{10}(<M_{dust}>/\mathrm{M_{\odot}}) = 6.54 \pm 0.19$, $<T_{dust}> = 17.47 \pm 0.97$\,K, and $\log_{10}(M_{dust}/M_{stars}) = -2.93 \pm 0.09$, respectively. The different cluster environments seem to have had little effect on the FIR properties of the galaxies and so we conclude that any environment dependent evolution, has taken place before the cluster was assembled.

\end{abstract}
\section{Introduction}

The Fornax cluster is a nearby example of a poor but relatively relaxed cluster. It has a recession velocity of 1379\,km\,s$^{-1}$ and a distance of 17.2\,Mpc, a mass of 7$ \times$10$^{13}$ M$_{\odot}$ and virial radius of 0.7\,Mpc~\citep{drinkwater01}. It is located away from the Galactic plane with a Galactic latitude of $-53.6^{\circ}$ in an area of relatively low Galactic cirrus. This makes it ideal for study at all wavelengths.

~\citet{drinkwater01} showed that despite Fornax's apparent state of relaxation, it still contains substructure, e.g. a small, in-falling group centred on NGC\,1316, 3$^{\circ}$ to the southwest. However, compared to the Virgo cluster, Fornax is very centrally concentrated and probably at a much later epoch of formation. This is also suggested by the strong morphological segregation that has taken place, leaving the cluster almost entirely composed of early-type galaxies.~\citet{drinkwater01} also noted that there exist two different populations, suggesting that while the giant galaxies are virialised, the dwarf population is still in-falling. Morphological segregation is not the only indicator of evolution in the cluster, the interstellar medium (ISM) of the galaxies also seems to have been affected by the cluster environment.~\citet{schroder01} found that 35 Fornax cluster galaxies were extremely \hi-deficient in comparison to a field sample. \hi\ is generally loosely bound to galaxies and as such is a good indicator of the effects of environmental processes. 

Dust, another constituent of the ISM, is also affected by the environment.~\citet{cortese10,cortese10b} showed that within a cluster like Virgo, dust can be stripped from the outskirts of a galaxy, truncating the dust disk. Dust is crucial for the lifecycle of a galaxy, as it allows atomic hydrogen to transform on its surface into molecular hydrogen and is thus essential for star formation. Around half the energy emitted from a galaxy is first emitted by stars, then reprocessed by dust, and re-emitted from 1\,$\mu m$ to 1\,mm~\citep{driver08}. Thus, to better understand the physical processes affecting galaxies it is crucial that we observe and understand the complete `stellar' spectral energy distribution (SED).

In contrast to Virgo, Fornax has only very weak X-ray emission~\citep{fornaxxray,virgoxray}, which traces the hot intra-cluster gas. Compared to Virgo, this lack of an intra-cluster medium (ICM) along with a lower velocity dispersion ($\sim$\,300\,km\,s$^{-1}$) reduces the efficiency of mechanisms such as ram pressure stripping. We can estimate the efficiency of ram pressure stripping using $E \propto t_{cross} \delta v^{2} \rho _{gas}$~\citep{gunn72}, where ($E$) is the stripping efficiency of a cluster, with a velocity dispersion ($\delta v$), central gas density ($\rho _{gas}$) and a crossing time $t_{cross}$. Both Virgo and Fornax have a similar crossing time, $t_{cross}\sim 10^{9}$\,yr which is much less than their relaxation time t$_{relax} \sim 10^{10}$\,yr~\citep{boselli06}. Virgo has a velocity dispersion which is $\sim$\,4$\times$ greater and an ICM $\sim$\,2$\times$ as dense as Fornax~\citep{chen07, boselli06}, indicating that Fornax may be $\sim$\,32$\times$ less efficient than Virgo in removing a galaxy's ISM via ram pressure stripping. 

Fornax's higher galaxy density, lower ICM density and lower velocity dispersion suggest that galaxy-galaxy tidal interactions will play a more important role than in a more massive cluster like Virgo~\citep{combes88,kenney95}. 

Whilst the near-infrared (NIR, 1 - 5\,$\micron$) and mid-infrared (MIR, 5 - 20\,$\micron$) emission from a galaxy is dominated by the old stellar population and complex molecular line emission, respectively, the far-infrared (FIR, 20 - 500\,$\micron$) and sub-mm regime (500 - 1000\,$\micron$) is dominated by dust emitting as a modified blackbody. Although there are a few small windows in the earth's atmosphere, most of the infrared spectrum is absorbed and is either impractical or not possible to observe from the ground, so the infrared wavelength regime is best studied from space-based observatories. 

In 1983, the $IRAS$~\citep{IRAS} (10\,-\,100\,$\mu$m) all-sky survey opened up the extragalactic infrared sky for the first time. Of particular interest to us is the first detection of FIR sources associated with the Fornax cluster.~\citet{wang91} found 5 $IRAS$ sources matching known Fornax galaxies inside the bounds of our survey and located preferentially towards the outskirts of the cluster. Since $IRAS$, very little further study has been undertaken of the Fornax cluster in the MIR or FIR. However, other cluster observations with $ISO$~\citep{ISO}(10\,-\,70\,$\mu$m) indicated that MIR emission, originating from hot dust ($\sim$60\,K) correlates well with \hii\ regions, implying that it is heated primarily by star formation (SF)~\citep{popescu02}. In contrast, FIR emission from cold dust ($\approx $20\,K) had a nonlinear correlation with H$_{\alpha}$ luminous regions, indicating a link to the older, more diffuse stellar population. Most significantly, they found a cold dust component that, in some cases, was less than 10\,K, though $ISO$ lacked the longer wavelength photometric coverage to constrain the Raleigh-Jeans blackbody tail of this `cold dust emission'. The $Spitzer$ Space Telescope~\citep{spitzer} (3\,-\,160\,$\mu$m) was launched in 2003. Using the MIPS instrument~\citet{edwards11} showed that in the Coma cluster SF is suppressed in the cluster and this suppression decreases with distance from the cluster core. All the instruments described above lacked photometric coverage at wavelengths needed to constrain the temperature and mass of cold dust ($T \textless 20$K). The $Herschel$ Space Observatory~\citep{pilbratt10} rectified this problem as it was able to survey large areas of sky at longer FIR wavelengths and with superior resolution and sensitivity.

The $Herschel$ Fornax Cluster Survey~\citep[HeFoCS;][]{davies12} observations discussed in this paper make use of the superior observational characteristics of the $Herschel$ Space observatory to address the problems highlighted above. This paper is one in a series of papers in which we compare the properties of galaxies in both the Virgo and Fornax clusters. Other papers in this series are: Paper I~\citep{davies10} examined the FIR properties of galaxies in Virgo cluster core; Paper II~\citep{cortese10} studied the truncation of dust disks in Virgo cluster galaxies; Paper III~\citep{clemens10} constrained the lifetime of dust in early-type galaxies; Paper IV~\citep{smith10} investigated the distribution of dust mass and temperature in Virgo's spirals; Paper V~\citep{grossi10} examined the FIR properties of Virgo's metal-poor, dwarf galaxies; Paper VI~\citep{baes10} a FIR view of M87; Paper VII~\citep{delooze10} detected dust in dwarf elliptical galaxies in the Virgo cluster; Paper VIII~\citep{davies12a} presented an analysis of the brightest FIR galaxies in the Virgo cluster; Paper IX~\citep{magrini11} examined the metallicity dependence of the molecular gas conversion factor; Paper X~\citep{corbelli12} investigated the effect of interactions on the dust in late-type Virgo galaxies; Paper XI~\citep{pappalardo12} studied the effect of environment on dust and molecular gas in Virgo's spiral galaxies; Paper XII~\citep{auld12} examined the FIR properties of an optically selected sample of Virgo cluster galaxies; Paper XIII~\citep{alighieri13} investigated the FIR properties of early-type galaxies in the Virgo cluster; Paper XIV~\citep{delooze13} studied Virgo's transition-type dwarfs and Paper XVI~\citep{Davies14} presented an analysis of metals, stars, and gas in the Virgo cluster. Six further papers~\citep{Boselli10,hrslate,boquien12,ciesla12,smith12,eales12} discuss the HeViCS galaxies along with other galaxies observed as part of the Herschel Reference Survey (HRS).

\section{Data and Flux Measurement}

This paper is based on the methodology of~\citet{auld12}. As in~\citet{auld12} we initially use an optical catalogue of cluster galaxies to select our targets and then measure the FIR flux density at those locations. \subsection{Optical Data}

\subsubsection{Fornax Cluster Catalogue }
The Fornax Cluster Catalogue~\citep[FCC;][]{ferguson90}, was created from visual inspection of photographic plates taken with the Du Pont 2.5m reflector at the Las Campanas Observatory. It is complete to m$_{BT}\sim 18$, and contains members down to m$_{BT}\sim 20$. Although this catalogue is 20 years old it is still the best optical catalogue available. It is equivalent to the Virgo Cluster Catalogue~\citep[VCC;][]{vcc} used by~\citet{auld12} and so enables a good comparison between the two clusters.~\citet{ferguson90} assigned cluster membership mainly based on morphology and the detail that could be observed in the images. There are now 104 radial velocity measurements of FCC galaxies which indicate that 6 of them are outside the cluster, (FCC\,97, 141, 189, 233, 257, and 287). These were removed from our sample. 

We use the FCC's positions, optical sizes and shapes as a starting point from which to fit an aperture and measure the FIR emission for each galaxy (see Figure~\ref{fig:by-eye}). 

\subsubsection{SuperCOSMOS Sky Survey}
Due to the angular resolution of $Herschel$ ($\sim$18" at 250\,$\micron$) many sources are either confused or blended with nearby or background galaxies. To help overcome this problem we have used the SuperCOSMOS Sky Survey (SSS)~\citet{hambly01}. The SSS's comparatively high angular resolution allows us to discern if a FIR source is likely to be associated with a Fornax cluster galaxy. We overlay FIR contours on SSS r-band images to determine whether a FIR source can be clearly associated with a single FCC galaxy.

\subsection{Herschel data}

\subsubsection{HeFoCS data}
\begin{figure}
\centering
\includegraphics[width=\linewidth]{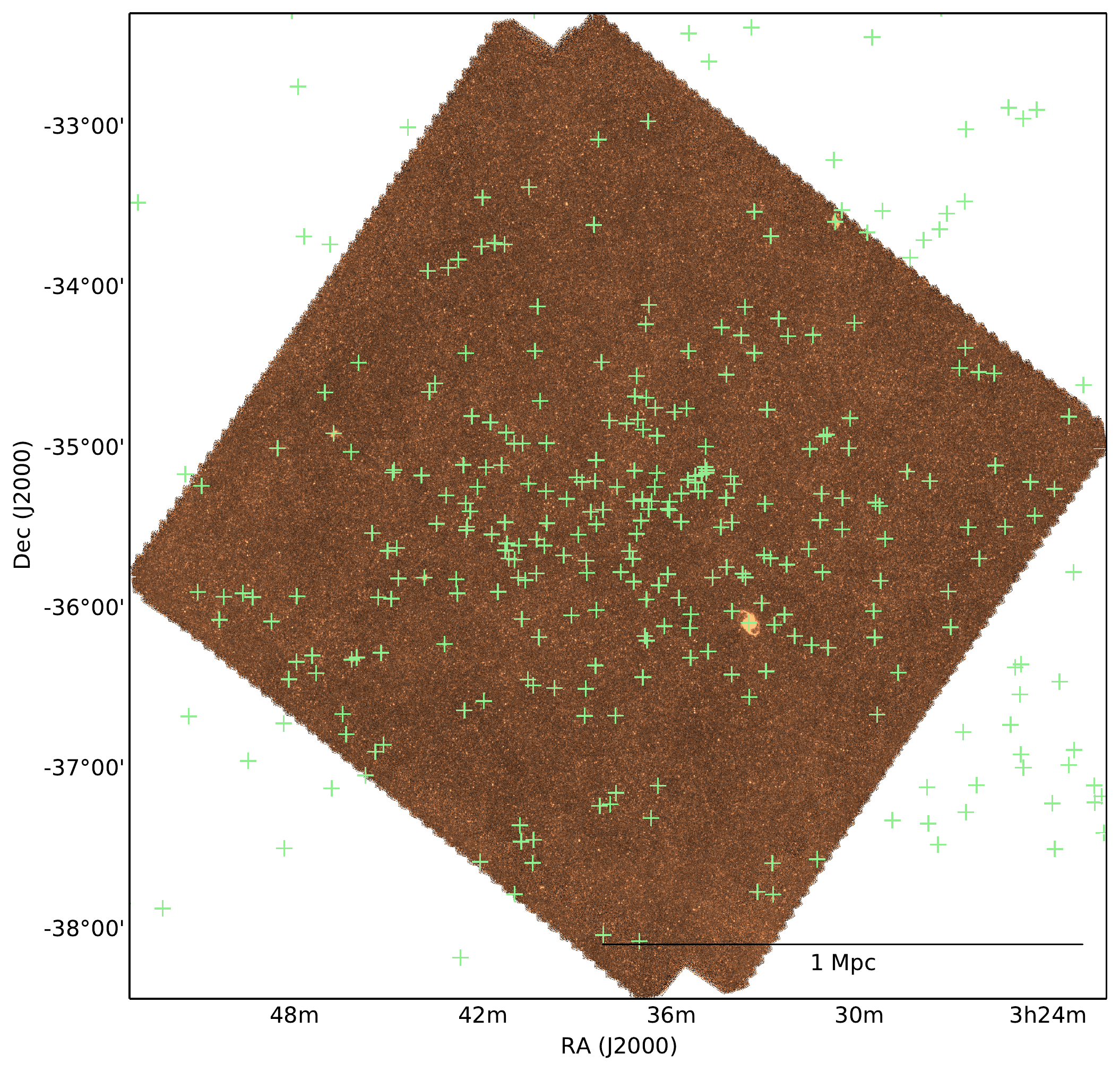}
\caption{The HeFoCS $250\micron$ image, with a green cross marking the position of every galaxy from the optical FCC catalogue. The $Herschel$ data miss galaxies in the outskirts of the cluster. A scale bar of 1\,Mpc is marked in the lower right hand corner, assuming a distance to the cluster of 17.2\,Mpc. }
\label{fig:fornaxplot}
\end{figure}

The HeFoCS observations cover a $4\,^{\circ}\,\times\,4\,^{\circ}$ tile centred on NGC\,1399 [$\alpha = 03^{h}38^{m}29.8^{s}, \delta = -35^{o}27'2.7"$], the central elliptical galaxy. This is an area of apparently low galactic cirrus when compared to the Virgo cluster field~\citep{davies12}. The region contains $\sim$\,70\,\% of the area covered by the FCC catalogue, Figure~\ref{fig:fornaxplot}. It should be noted that there is an unavoidable misalignment between SPIRE and PACS due to their respective locations on the $Herschel$ focal plane. This misalignment leads to a loss of 30 galaxies that are not in the PACS maps, and are only observable by SPIRE. In total 60\,\% of the FCC cluster galaxies are observed in all 5 bands (100, 160, 250, 350 \& 500\,$\micron$). 

The FIR maps in this paper are identical to those used in~\citet{davies12} and a full description of the data reduction for the HeFoCS data is available in that paper. Briefly, the HeFoCS observations are taken using PACS ($100\,\&\,160\micron$)~\citep{poglitsch10} and SPIRE ($250\, , 350\,, \& \,500\micron$)~\citep{griffin10} in parallel mode with a fast scan rate (60\,arcsec\,s$^{-1}$), and our final maps consist of 4 scans ($2\times2$ orthogonal cross-linking scans).

PACS data were taken from level 0 to level 1 using the standard pipeline, then the 4 scans were combined with the \textit{Scanamorphous} map maker~\citep{roussel13}. 

SPIRE data were processed with a customised pipeline from level 0 to level 1, which is very similar to the official pipeline. The difference being the use of a method called BriGAdE (Smith et al., in preparation), in place of the standard \textit{temperatureDriftCorrection}. BriGAdE effectively corrects all the bolometers for thermal drift without removing large extended structures like Galactic cirrus. These scans are then combined using the na\"{i}ve mapper in the standard pipeline. 

The final maps have pixel sizes of 2, 3, 6, 8 and 12 arc seconds and $1\sigma$ noise over the entire image of 0.5, 0.7, 0.7, 0.8 and 0.9\,mJy\,pixel$^{-1}$ (or 9.9, 9.2, 8.9, 9.4 and 10.2 mJy\,beam$^{-1}$ ) for 100, 160, 250, 350 and 500\,$\micron$, respectively\footnote{In this paper we present updated values for the global noise. These are lower than those presented by~\citet{davies12}. Our method for calculating the gobal noise is described in Section~\ref{sec:comp}.}. 
The approximate Full Width Half Maximum (FWHM) of the $Herschel$ beam is 11, 14, 18, 25 \& 36 arc seconds, at 100, 160, 250, 350 \& 500\,$\micron$, respectively. At the distance of Fornax 10 arc seconds\,$\simeq$\,1\, kpc giving us the potential to resolve many Fornax galaxies. For example, the three biggest galaxies in the cluster are NGC\,1365, 1399, and 1380 with optical diameters of 5.5, 3.8, and 2.7 arc minutes, respectively. 

\subsubsection{Comparsion with HeViCS}
\label{sec:comp}

As much of this paper is based on a comparison between Fornax and Virgo, it is therefore worthwhile to examine the difference between the HeFoCS and HeViCS data. The FIR maps of both surveys are created using identical data reduction techniques. However, they differ with respect to depth and spatial coverage of the clusters.  

First we consider the depth of the surveys. The HeViCS maps consist of 8 scans (4\,x\,4 orthogonal cross-linking scans), twice as many as the HeFoCS maps, leading to a $\sim$\,$\sqrt{2}$ reduction of instrumental noise.~\citet{auld12} calculated instrumental and confusion noise, showing that the HeViCS SPIRE bands were effectively confusion noise limited (70\,\% of the overall noise is from the confusion noise at 250\,$\micron$). Consequently, when planning the HeFoCS we requested 4 scans as this offered almost confusion limited maps with half the time required for a single HeViCS tile.

In order to asses the ratio of the global noise in the HeViCS and HeFoCS maps, we measure the pixel-pixel fluctuations and apply an iterative 3$\sigma$ clip to remove bright sources. The global noise in the HeViCS and HeFoCS at 250\,$\micron$ is thus, 7.5 and 8.9 mJy\,beam$^{-1}$, respectively, yielding a ratio between the two of 1.19. This ratio is half as much as one would expect from a simple $\sqrt{2}$ increase in depth if the maps were purely instrumental noise limited, thus showing that the surveys are reasonably well suited for comparison.

Second we consider the coverage of the HeViCS and HeFoCS FIR maps of their respective clusters. This is not a straightforward task, the clusters have very different physical sizes and states of relaxation - Virgo is far more `clumpy' than Fornax. The irregular shape of Virgo lead to the HeViCS FIR maps comprising of 4 tiles ($4\,^{\circ}\,\times\,4\,^{\circ}$) running North to South, whereas the HeFoCS is only a single tile ($4\,^{\circ}\,\times\,4\,^{\circ}$). A possible solution is to use the fraction of the VCC and FCC galaxies that lay inside the boundary of each FIR survey, this is incidentally $\frac{2}{3}$ for both, showing again that the HeViCS and HeFoCS are well suited for a FIR comparison of the two clusters. 

\subsubsection{Missing FIR sources}
\begin{figure}
\centering
\includegraphics[trim = 0 0 0 10,width=\linewidth]{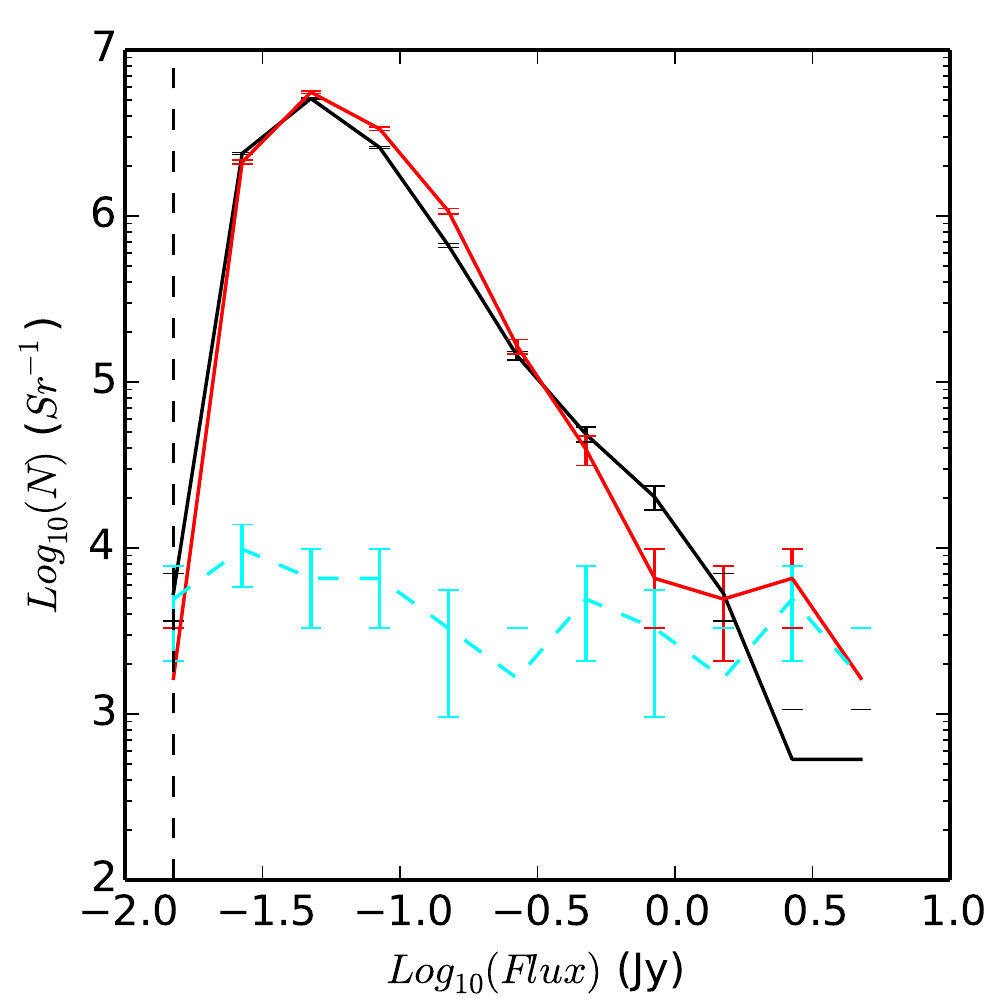}
\caption{A histogram of the 250 $\micron$ flux density of galaxies against number per steradian. The black and red lines are the NGP (from H-ATLAS) and the HeFoCS fields respectively. The cyan dashed line represents the detected FCC galaxies. The errors are simply $\sqrt{N}$. The vertical dashed line marks the minimum flux density detectable ($\sim$\,15\,mJy) given the minimum detection area and 1.6\,$\sigma$ noise level in the HeFoCS data. }
\label{fig:numbercounts}
\end{figure}

By using an optical catalogue we run the risk of missing a population of FIR sources not detected in the optical. Conversely we could make our selection in the FIR, but then there is no way of determining which sources are in the cluster.

Below, we show that a large population of cluster FIR sources without optical counterparts does not exist. We do this by comparing the number counts of sources in the HeFoCS 250\,$\micron$ map with data extracted from the H-ATLAS~\citep{eales10} North Galactic Pole (NGP) field, which has no foreground cluster. By comparing histograms of number counts in these two fields, one `looking' through the Fornax cluster and the other a purely background reference, we can look for evidence of a FIR excess of sources in the cluster. 

We only use the 250\,$\micron$ band to do this, as we impose a selection in our FIR catalogue such that each galaxy must be detected at 250\,$\micron$ (see below). The NGP data and data reduction are fully described in Valiante et al. (in preparation). Here we provide a brief description. The data consists of a scan and cross scan, which have been reduced using the same pipeline as the HeFoCS data. The only difference is that the NGP 250\,$\micron$ map is gridded onto 5" pixels, whereas the HeFoCS data are gridded onto 6" pixels. Using the same iterative 3$\sigma$ clipping method as described above, the global noise in the NGP map at 250\,$\micron$ is 13.0\,mJy\,beam$^{-1}$, a factor of 1.5 greater than the HeFoCS data in the same band. The NGP covers a $\sim$\,180 square degree area of sky, from which we have extracted a 20 square degree area in the north east of the map. This selection is used to avoid the Coma cluster (D\,=\,100\,Mpc) which is located in the south west. 

We use the software, \textit{SExtractor} to measure the flux density of all the sources in each map, taking care to use an identical method. \textit{SExtractor} `grids up' each map and calculates the noise in each sub-grid. The parameter that controls this is `meshsize', we fix at 100 arcmin$^2$ as this is much greater than the size of any of our foreground galaxies. The detection threshold was set at 1$\sigma$ and 1.6$\sigma$ above the local background for the NGP and HeFoCS maps, respectively. These different threshold values are used to ensure comparable sensitivity in each field given the smaller pixel sizes and higher noise per pixel in the NGP field. Another requirement was that the detection size of a source was greater than the SPIRE beam area at 250\,$\micron$  (450 arcsec$^2$).  

Figure~\ref{fig:numbercounts} shows the number counts generated from using the above approach. The black and red lines represent the NGP and HeFoCS fields, respectively. The vertical dashed line marks the minimum flux density detectable ($\sim$\,15\,mJy) given the minimum detection area and 1.6\,$\sigma$ noise level in the HeFoCS data. The cyan dashed line is for our Fornax FIR catalogue, as presented in this paper (see below). The black and red lines trace each other very well within the $\sqrt{N}$ errors below about 1\,Jy, brighter than this there is a small excess due to the presence of Fornax cluster galaxies. In conclusion we find no evidence for a significant excess population of FIR sources that are not associated with the optical sources in the FCC. There are too few 250$\micron$ detections in Fornax to create a statistically meaningful luminosity function (about 3 galaxies per bin in Figure~\ref{fig:numbercounts}), however, interestingly it has a similar flat luminosity function as found by~\citet{Davies14} for the Virgo cluster.

\subsection{FIR flux density measurements}
\label{sec:sourcemeasurement}

\subsubsection{General approach}
\begin{figure*}
\centering
\includegraphics[trim = 35 0 45 15, clip,width=\linewidth]{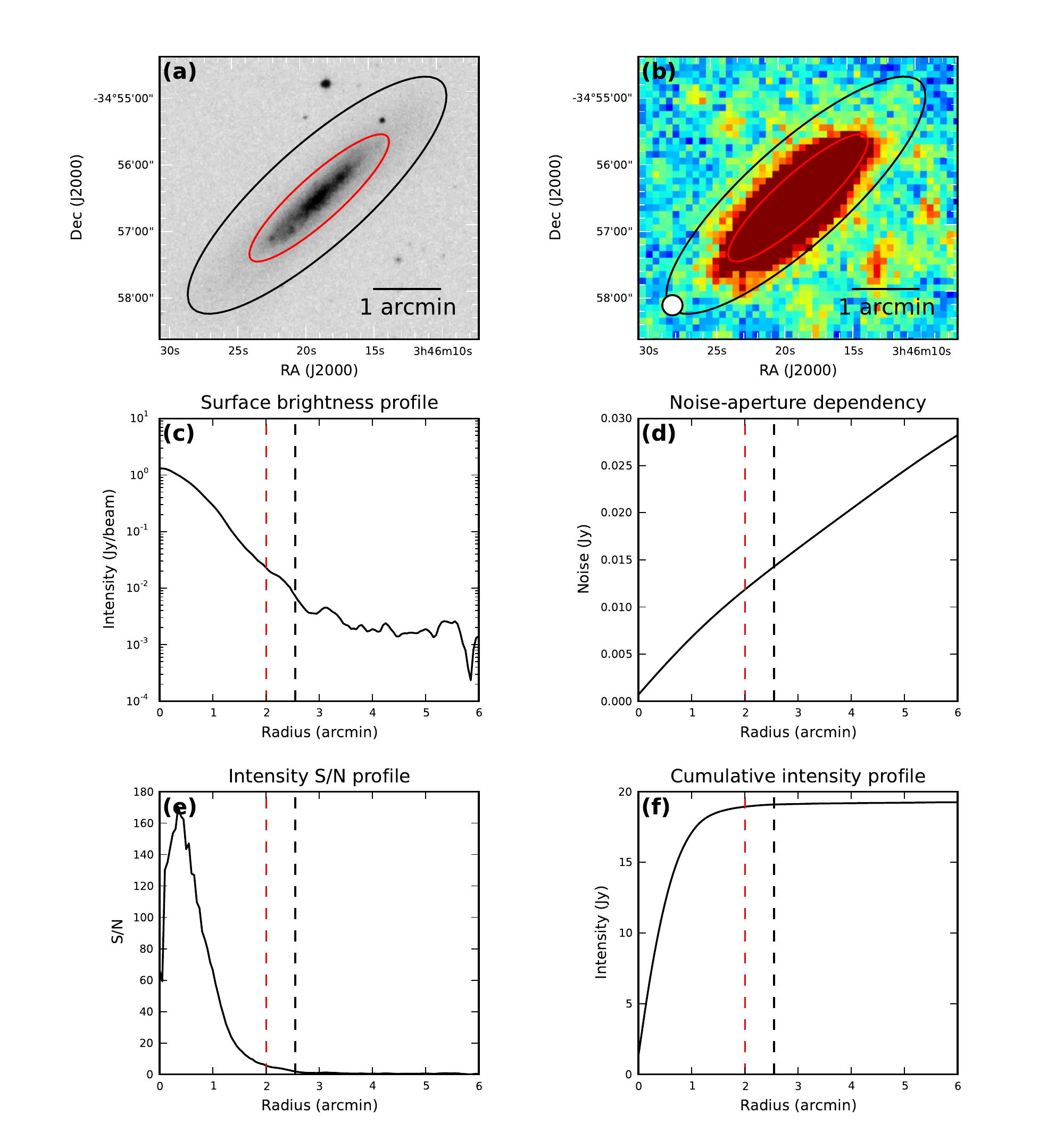}
\caption{This is the postscript output generated for FCC312. Excluding the upper lefthand panel, all panels refer to the $Herchel$ 250\,$\micron$ band. (a) The superCOSMOS r-band optical image, the red and black ellipses show the optical D$_{25}$ and the FIR extent of the galaxy (see text for definition). (b) The raw sub-image cutout of the HeFoCS map. The beam size is shown in the lower left hand corner. (c) The surface brightness profile. (d) Noise for an equivalent sized circular aperture (see text for definition). (e) S/N per annuli. This shows the cut off when S/N$\le$2. (f) A cumulative intensity profile. The red and black dashed lines show the optical and FIR extent, respectively.}
\label{fig:fccoutput}
\end{figure*}

We have used a semi-automated source measurement program written in IDL, to measure the FIR flux density of each galaxy. This program is fully described and extensively tested by~\citet{auld12}. The method is briefly described below. 

Of the FCC galaxies, 237 fall into the SPIRE maps and 201 fall into both the PACS and SPIRE maps. The optical parameters (position, eccentricity, optical diameter $D_{25}$ and position angle\footnote{Position angle and eccentricity are not listed in the FCC but were obtained using the online database \textit{Hyperleda},~\citep{hyperleda}.}) from the FCC were used to make an initial estimate of the shape and size of the FIR emission. Previous studies~\citep{cortese10, pohlen10} as well as the equivalent to this study in the Coma cluster (Fuller et al., in preparation), show that FIR emission is well traced by the optical parameters of late-type galaxies. Whereas early-type galaxies typically show more compact dust emission~\citep{smith12}. The optical parameters are only used to make an initial estimate for creating masks. The program then iterates, to create masks and apertures that best match the diameter and ellipticity of the FIR emission. For the following explanation, it may serve the reader to consult Figure~\ref{fig:fccoutput}. 

The flux measurement process starts by extracting a 200\,$\times$\,200 pixel sub-image from the raw map as shown in Figure~\ref{fig:fccoutput}b. To measure the background of the sub-image, all nearby galaxies including the galaxy being measured are initially masked at 1.5\,$\times$\,D$_{25}$. If the optical extent of the galaxy is such that this sub-image is not large enough to give an accurate background estimation, then the program will increase the size of the sub-image, up to 600$\times$600 pixels for SPIRE and 1200$\times$1200 for PACS. 

The background estimation has to deal with the near confusion limited SPIRE maps and instrumental noise in the PACS maps. This program was originally written for use in the HeViCS maps where galactic cirrus was also a major problem. In order to remove bright background galaxies and galactic cirrus~\citet{auld12} used a 98\% flux clip and then fitted the remaining pixels with a 2D polynomial. The flux clip, removes bright background galaxies by masking out the brightest 2\% of pixels, this ensures that the 2D polynomial is only fitting the galactic cirrus. Cirrus is not obviously present in the HeFoCS maps and as such the 98\% clip has been retained and then the median pixel value of the masked sub-image taken as the background value.  

\begin{figure}
\centering
\includegraphics[trim = 4 48 0 25, clip,width=\linewidth]{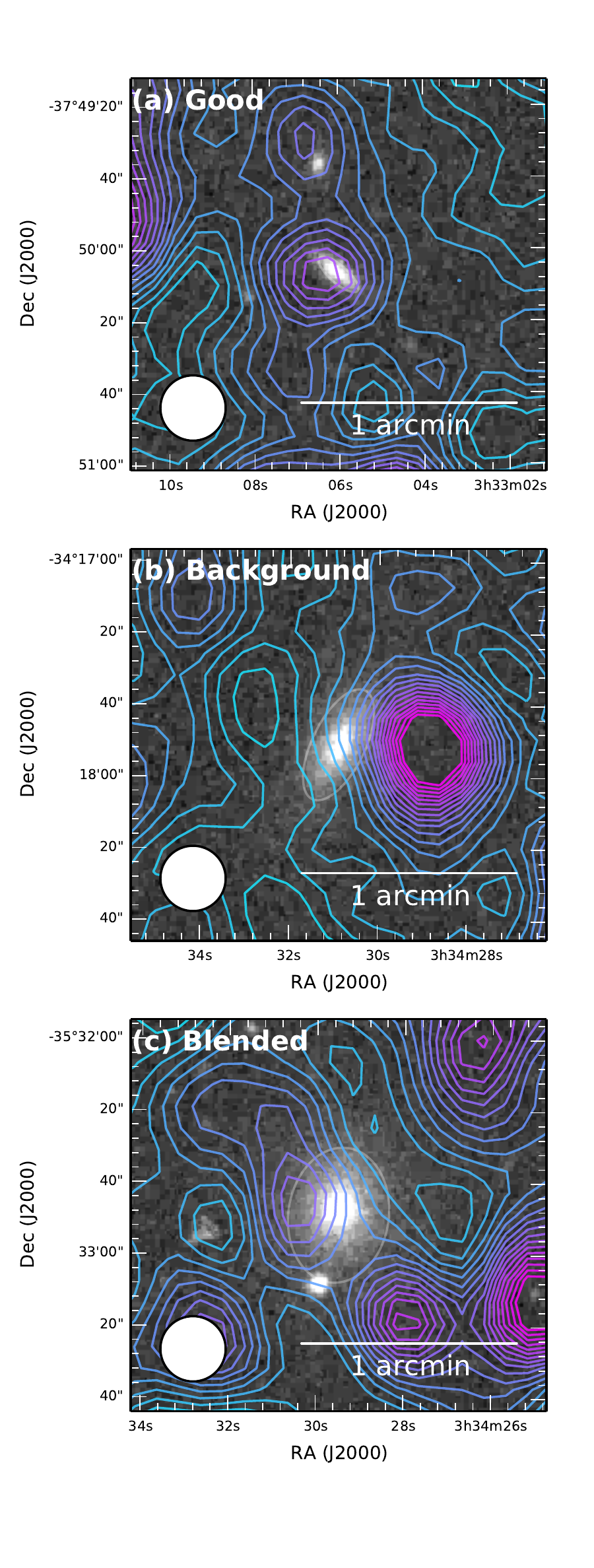}
\caption{The SPIRE $250\micron$ contour map, plotted over the superCOSMOS r-band image of FCC\,117, 135 and 136 for sub-figures (a), (b) and (c) respectively. The beam size is shown in the lower left hand corner. The white ellipses indicate the optical ($D_{25}$) extent of each galaxy. (a) a galaxy that by-eye we flagged as a good detection as it is coincident with the FIR contours. (b) this galaxy was removed as it is clearly a bright background source that does not appear in the optical image. (c) the FIR source cannot be uniquely identified, it looks to be comprised of more than one source and as such was removed.}
\label{fig:by-eye}
\end{figure}

 We measured total flux, surface brightness, aperture noise (fully described in Section~\ref{sec:noise}) and signal to noise (S/N) along annuli of increasing radius centred on the galaxy optical centre. The shape of the annuli is based on the galaxy optical parameters convolved with the appropriate point spread function (PSF). We plot the corresponding radial profiles in Figure~\ref{fig:fccoutput} c, e \& f, respectively. The FIR diameter\footnote{As some galaxies are not resolved D$_{FIR}$ in some cases be defined by the PSF of the $Herschel$ beam and will not be representative of the extend of dust in the galaxy.} D$_{FIR}$ is defined where the S/N profile drops below 2. This D$_{FIR}$ is used to replace the 1.5$\times D_{25}$ used to make the initial mask. The process iterates until the mask and the D$_{FIR}$ value converge. Only then were aperture corrections applied according to~\citet{ibar10} and Griffin \& North (In preparation). The aperture corrections take into account the encircled energy fraction within the chosen aperture size. The median aperture corrections are 1.00, 1.00, 0.83, 0.90 and 0.92 at 100, 160, 250, 350 \& 500\,$\micron$, respectively. 

If the total S/N value was less than 3, the sub-image was then searched optimally for a point source. After convolving with the relevant PSF, the maximum value within the FWHM of the PSF centred on the optical position was taken as the flux. The noise was calculated according to~\citet{marsden09} and~\citet{chapin11}, which involved plotting a histogram of all the pixels in the PSF-convolved sub-image and fitting a Gaussian function to the negative tail. The FWHM of this Gaussian is then used to estimate the combined instrumental and confusion noise. This has been summed in quadrature with the calibration uncertainty (see below) to obtain a value for the total noise. If the S/N was still less than 3 we consider the object undetected and set an upper limit on the flux equal to 3 times the noise in the PSF-convolved sub-image. This marked the end of the automatic source measurement process. The output is in the form of postscript files for each galaxy, as shown in Figure~\ref{fig:fccoutput}.  

\subsubsection{Dealing with blending and contamination}

$Herschel's$ comparatively large FWHM can lead to unavoidable contamination by FIR background sources, which could be falsely identified as Fornax galaxies. The level of this contamination is estimated in Section~\ref{sec:contaminations}. We have plotted the 250\,$\micron$ map as contours over a superCOSMOS image of each galaxy and its immediate environment. As shown in Figure~\ref{fig:by-eye}, if a galaxy could not be clearly separated from a nearby or background galaxy we removed it from our catalogue. Figure~\ref{fig:by-eye}a shows a FIR source that is clearly coincident with a Fornax galaxy. Figure~\ref{fig:by-eye}b shows a background source that is brighter than the 3$\sigma$ noise limit and has been registered as a detection by our program. Figure~\ref{fig:by-eye}c may be a detection, however, we cannot separate it from another apparent detection, so it was also removed. For galaxies that have been eliminated from our final catalogue through this process we set an upper limit on their flux density equal to the 3$\sigma$ noise from the PSF-convolved map. As in~\citet{auld12} we impose a strict criterion that a galaxy must be detected at 250\,$\micron$ as this provides the best combination of sensitivity and resolution (see below).

\begin{figure}
\centering
\includegraphics[width=\linewidth]{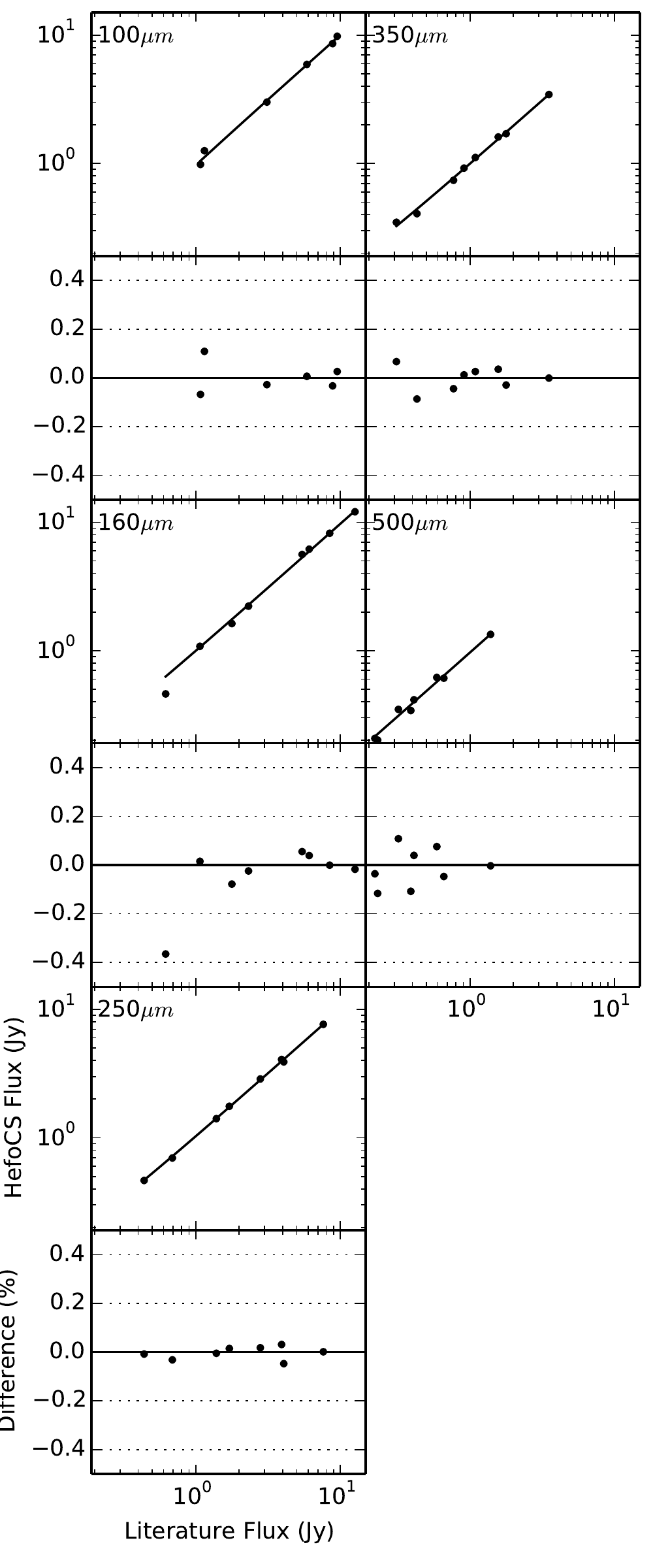}
\caption{The HeFoCS fluxes plotted against the Davies et al. (2013) values for the bright galaxy sample. The residual plot below shows the percentage deviation from the fitted line.}
\label{fig:fluxv}
\end{figure}
\subsubsection{Total uncertainty estimate}
\label{sec:noise}

The total uncertainty is estimated from the calibration uncertainty, $\sigma_{cal}$ and aperture uncertainty, $\sigma_{aper}$, summed in quadrature.

For SPIRE, $\sigma_{cal}$ is based on single scans of Neptune and on an assumed model of its emission. The final error for each band is estimated to include 4\% correlated and 1.5\% from random variation in repeated measurements, as well as 4\,\% due to uncertainty in the beam area. The SPIRE observer's manual\footnote{http://herschel.esac.esa.int/Docs/SPIRE/html/spire\_om.html} suggests that these should be added together, leading to a SPIRE $\sigma_{cal}$ of 9.5\%. 

For PACS, $\sigma_{cal}$ is based on multiple sources with different models of emission. The PACS observer's manual\footnote{http://herschel.esac.esa.int/Docs/PACS/html/pacs\_om.html} lists the uncorrelated uncertainties as 3\,\% \& 4\,\% for 100 and 160\,$\micron$, respectively, and the correlated uncertainty is given for point sources as 2.2\,\%. However, the data used for calculating these uncertainties were reduced and analysed in a different way than the HeViCS and HeFoCS PACS data. Here we use the same value for total error as in~\citet{auld12}, i.e. 12\,\%.

To calculate the aperture uncertainty ($\sigma_{aper}$) a large number of apertures of a fixed size were placed randomly on each sub-image. We measure the total flux in each aperture, then by applying an iterative 3$\sigma$ clipping procedure use $\sigma$ as the uncertainty for that size of aperture. Repeating this for a range of aperture sizes allows us to estimate the aperture uncertainty as a function of size~\citep{ibar10}. This method takes into account, both confusion noise and instrumental noise. Figure~\ref{fig:fccoutput}d shows such a plot of aperture uncertainty against radial distance for FCC312.~\citet{auld12} tested this method over an entire 4$^{\circ}$\,x\,4$^{\circ}$ tile in the southern region of Virgo and compared it to the results obtained on the sub-images. They found very good agreement between the two, within the typical radii of FIR emission for Virgo galaxies. At larger radii this relationship broke down, this was attributed to large scale structure in the HeViCS maps. 

\subsection{Flux verification}

\begin{table}
\centering
  \begin{tabular}{cccc}

  \hline
  Band $\micron$ & Gradient, M & Intercept, C \\ \hline
  100      & 0.992\,$\pm$\,0.025    & 0.007\,$\pm$\,0.151   \\
  160      & 0.938\,$\pm$\,0.018    & 0.054\,$\pm$\,0.113   \\
  250      & 1.015\,$\pm$\,0.015    & -0.011\,$\pm$\,0.055  \\
  350      & 1.024\,$\pm$\,0.015    & -0.001\,$\pm$\,0.024   \\
  500      & 0.963\,$\pm$\,0.033    & 0.003\,$\pm$\,0.021    \\
\hline
  \end{tabular}
  \caption{The parameters of the straight line fit shown in Figure~\ref{fig:fluxv}.}
  \label{tab:fluxv}
\end{table}

As a verification of our automated process we have compared our fluxes with the Fornax Bright Galaxy Sample (BGS)~\citep{davies12}, as shown in Figure~\ref{fig:fluxv}, and tabulated the gradients and intercepts in Table~\ref{tab:fluxv}.~\citet{davies12} matched 10 galaxies with IRAS~\citep{IRASPSC} and 5 with PLANCK~\citep{plancksc} sources finding good agreement in both cases. Table~\ref{tab:fluxv} shows overall that the results are consistent with a gradient of 1 and an intercept of 0.

\section{Analysis \& Modeling}

\subsection{SED fitting}
\label{sec:sed-1}
We have fitted a modified blackbody to every galaxy detected in at least 3 $Herschel$ bands (22 galaxies) in order to estimate dust mass and temperature. The fit is based on the equation: 

\begin{equation}
 S_{\lambda } = \frac{\kappa_{abs} M_{dust} B(\lambda , T_{dust})}{D^{2}},  \nonumber
\end{equation}

\noindent where $S_\lambda$ is the flux density , $M_{dust}$ is the dust mass, $T_{dust}$ is the dust temperature, $B(\lambda , T_{dust})$ is the Planck function, D is the distance($D_{Fornax}$\,=\,17.2\,Mpc) and $\kappa _{abs}$ is the dust absorption coefficient. The latter follows a power law modified by an emissivity ($\beta$), such that:

\begin{equation}
 \kappa_{abs} =  \kappa_{abs}(\lambda_{0}) \times \left ( \frac{\lambda_{0}}{\lambda}\right ) ^{\beta} \nonumber
\end{equation}

We assume that emission at these wavelengths is purely thermal and from dust at a single temperature with a fixed $\beta = 2$ emissivity. We use $\kappa_{abs}(350 \micron)$ = 0.192\,m$^{2}$\,kg$^{-1}$ according to~\citet{draine03}. 

Although this is most likely an overly simplistic analysis, this approach has been used in previous works~\citep{davies10, davies12, smith12, auld12, Verstappen13} and shown to fit the data very well in the FIR/sub-mm regime.~\citet{bianchi13} showed that using a single component modified blackbody returns equivalent results to more complex models such as~\citet{draine07}. It is intended that in future papers we will explore two component fits as well as a variable beta emissivity. Derived dust masses and temperatures are given in Table~\ref{tab:FIRmasses} and the SED of each galaxy is shown in Figure~\ref{fig:sed_fits}. 

\subsection{Dust mass estimation}
\label{sec:250umtodust}

\begin{figure}
\centering
\includegraphics[trim= 0 0 0 0, width=\linewidth]{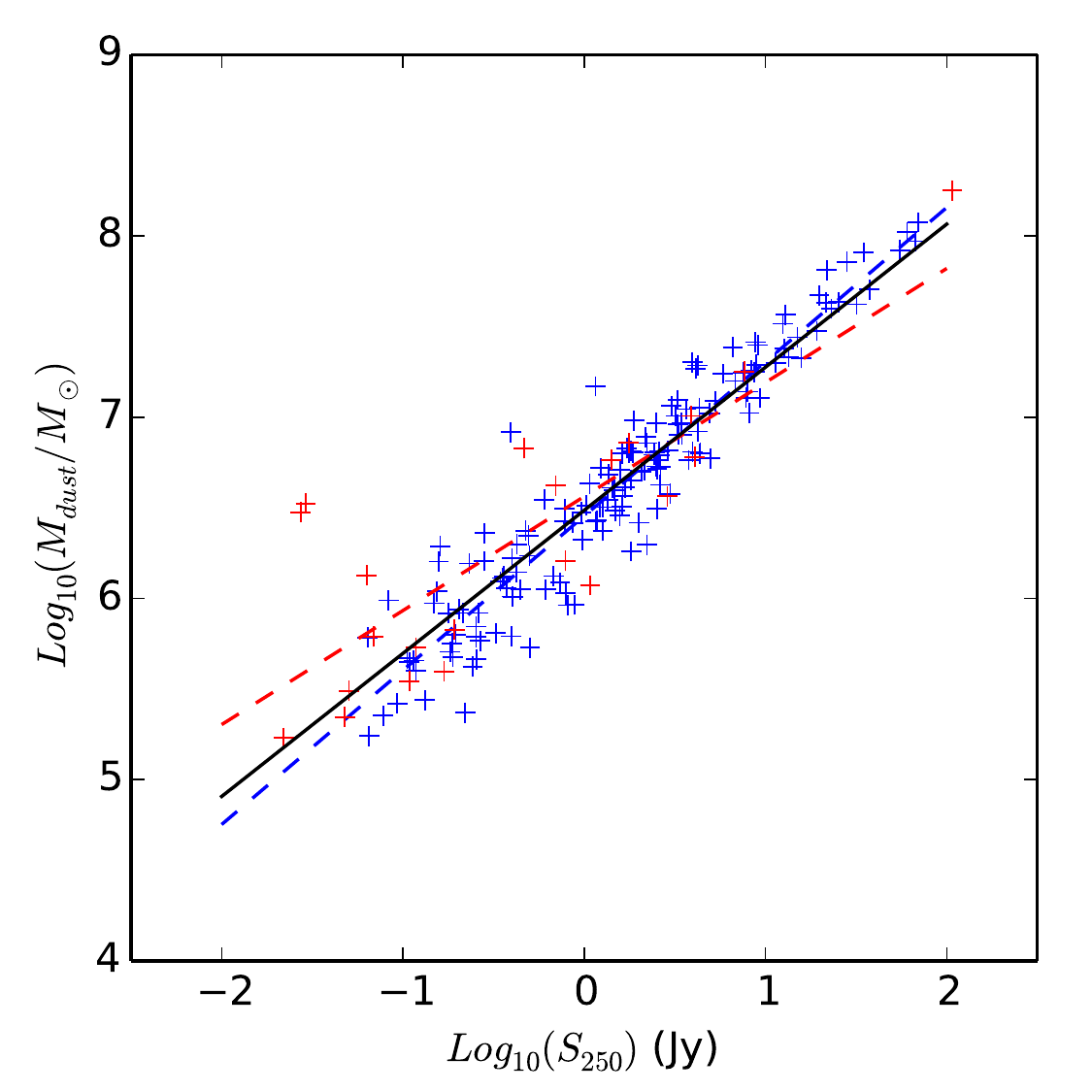}
\caption{Dust mass against 250$\micron$ flux density. The blue and red markers represent Virgo and Fornax galaxies, respectively. Dust masses are calculated from single temperature component, $\beta = 2$ emissivity modified blackbodies. The blue and red dashed lines represent the best fit to these for Virgo and Fornax, respectively. The black solid line represents the best fit to all the galaxies irrespective of which cluster they belong to.}
\label{fig:250todmass}
\end{figure}

Fornax has far less FIR detections than Virgo the dust mass can be calculated for only 22 of them through the SED fitting technique described above. We have performed much of the analysis in this paper with this sample of 22 Fornax galaxies. However, to best exploit the FIR data we have also used the 250$\micron$ flux density ($S_{250}$) as a proxy for dust mass, this added 9 galaxies to our Fornax FIR sample. Furthermore, this allows us to estimate an upper limit on dust mass for other galaxies not detected at 250$\micron$. 

In order to derive a relation between $S_{250}$ and $M_{Dust}$, we use galaxies from both Virgo and Fornax. Fornax galaxies are all assumed to lie at 17\,Mpc. For galaxies in Virgo we take $S_{250}$ and $M_{Dust}$ from~\citep{auld12} -- who used the same SED fitting method described above -- and scale the flux values to the distance of Fornax. 

In Figure~\ref{fig:250todmass} we plot $S_{250}$ against $M_{Dust}$ and fitted a single relation to all galaxies irrespective of which cluster they originate from:

\begin{equation}
 \log_{10}\left(\frac{M_{Dust}}{\mathrm{M_{\odot}}}\right) = 0.789 \times \log_{10}\left (\frac {S_{250}} {\mathrm{Jy}} \right) + 6.486  \nonumber
\end{equation}

\noindent We also fit relations to Virgo and Fornax individually and show them with a blue and a red dashed line, respectively. These have the same slope and incept within $1 \sigma$. Furthermore, we find a small range of dust temperatures. Consequently, we use a single relation for both clusters.

\subsection{Stellar masses}

\begin{figure}
\centering
\includegraphics[trim= 10 10 10 0, width=\linewidth]{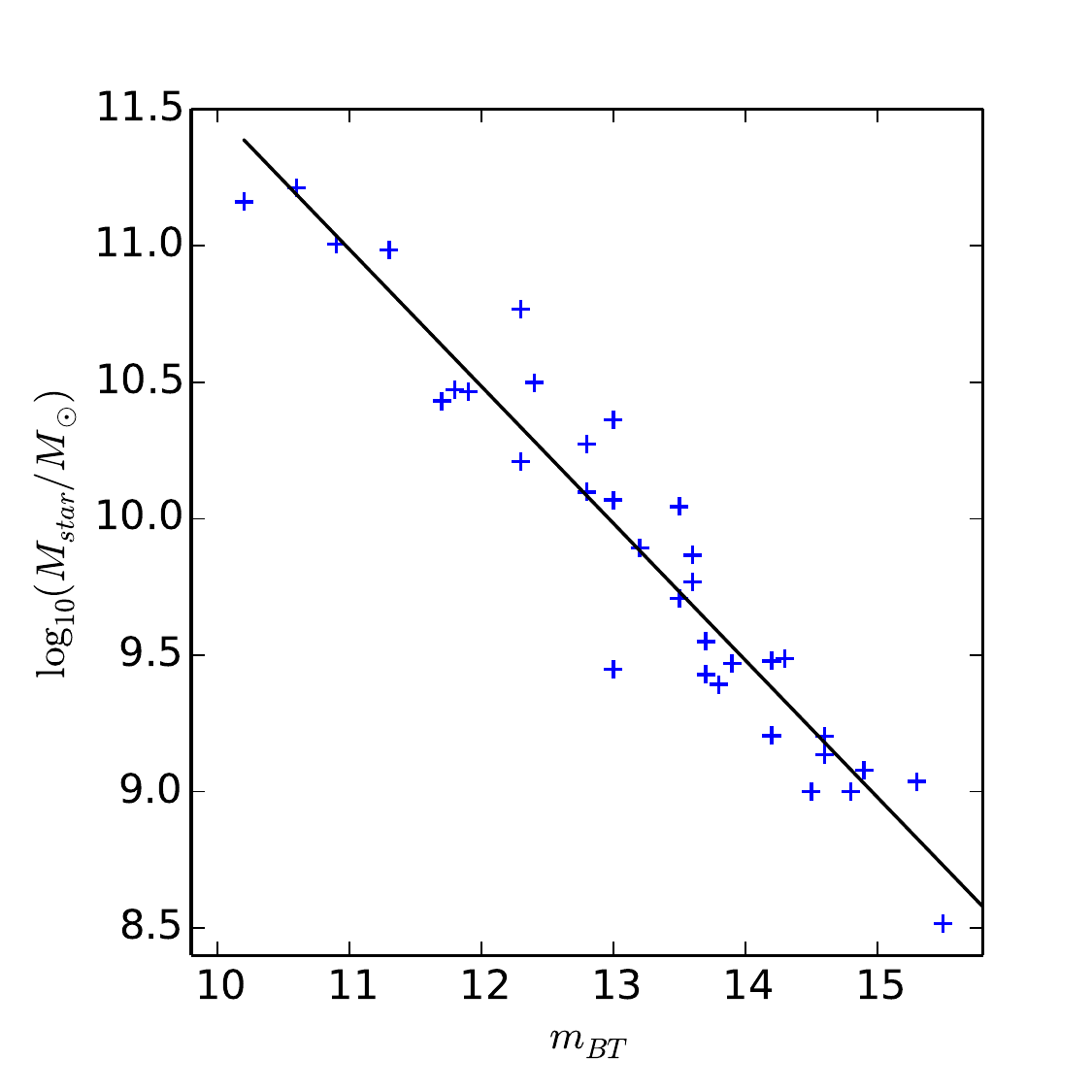}
\caption{Calculated stellar mass plotted against total blue magnitude. The blue points are galaxies with stellar masses calculated from their (B-V) colour and K band fluxes, the black line is a best fit line to these data.}
\label{fig:stellarbt}
\end{figure}

Only 35 galaxies in the FCC have both a (B-V) colour and K-band flux listed in~\textit{Hyperleda} and we have used this to calculate stellar masses using the prescription of ~\citet{bell03}:

\begin{equation}
 \log_{10}\left(\frac{M_{Star}}{\mathrm{M_{\odot}}}\right) = - 0.206 + 0.135(B-V) + \log_{10} \left ( \frac{L_{K}}{\mathrm{L_{\odot}}} \right) \nonumber
\end{equation}

\noindent Based on these 35 galaxies we find the following best-fitting linear relation between $m_{BT}$ and stellar mass:

\begin{equation}
 \log_{10}\left(\frac{M_{Star}}{\mathrm{M_{\odot}}}\right) = -0.51\,m_{BT} + 16.6. \nonumber
\end{equation}

\noindent We use this relation and the $m_{BT}$ value listed in~\textit{Hyperleda} to estimate the stellar mass of all remaining galaxies in the sample.

\subsection{Possible background contamination}
\label{sec:contaminations}
\begin{table}
\begin{tabular}{cccccc} 
\hline
Flux bin & N$_{a}$ & N$_{r}$ & N$_{p}$ & contamination & N$_{c}$ \\ 
(mJy) & ~ & ~ & (deg$^{-2}$) & \% & ~ \\ \hline 

20-45&9&17&516&4.14&9\\
45-100&5&1&297&2.38&5\\
100+ &3&1&59&0.48&1\\ 
\hline
\end{tabular} 
  \caption{Estimates of the contamination from background galaxies in the 250\,$\micron$ SPIRE band. N$_{a}$ is the number of sources accepted in each flux bin, N$_{r}$ is the number of sources that were rejected from the catalogue in each flux bin. $N_{p}$ is the source number density (see text). Contamination is the expected percentage contamination based on source counts. $N_{c}$ is the expected number of spurious contaminating sources.}
  \label{tab:con}
\end{table} 

In this section we try and assess whether our source rejection process has been reasonable given the background source counts. We assume that if extended FIR emission is found coincident with a Fornax galaxy it is reliable, and thus only concern ourselves with the point source population. 

We assume that the background sources are distributed randomly and uniformly across the sky with no cosmic variance. The number of contaminating sources is estimated using the number counts from the HeFoCS data (as described in Section 2.1) and then calculating the probability of a chance alignment with the 250\,$\micron$ SPIRE beam. We limit this analysis to the 250\,$\micron$ SPIRE band, as this was the band in which we made our by-eye inspection. It should also be noted that while the SPIRE bands are near confusion noise limited, the PACS bands are limited by instrumental noise. Consequently, PACS fluxes are far less likely contaminated by a background source. 

The contamination has been calculated within various flux intervals, as shown in Table~\ref{tab:con}. If done correctly we would expect the number of rejected galaxies to be roughly equivalent to the number of expected contaminating sources within the sum of the total area of apertures used. Table~\ref{tab:con} clearly shows that we have been over zealous in our rejection of sources in the 20\,-\,45\,mJy bin, however, in the 45\,-\,100\,mJy bin we have not rejected as many contaminating sources as the number counts predict. Overall we accept 17, reject 19 and estimate there are 15 contaminating galaxies at 250\,$\micron$. If we assume Poisson root N errors, then these small numbers are within $3 \sigma$. 

\section{Results}

\begin{figure*}
\centering
\includegraphics[width=160mm]{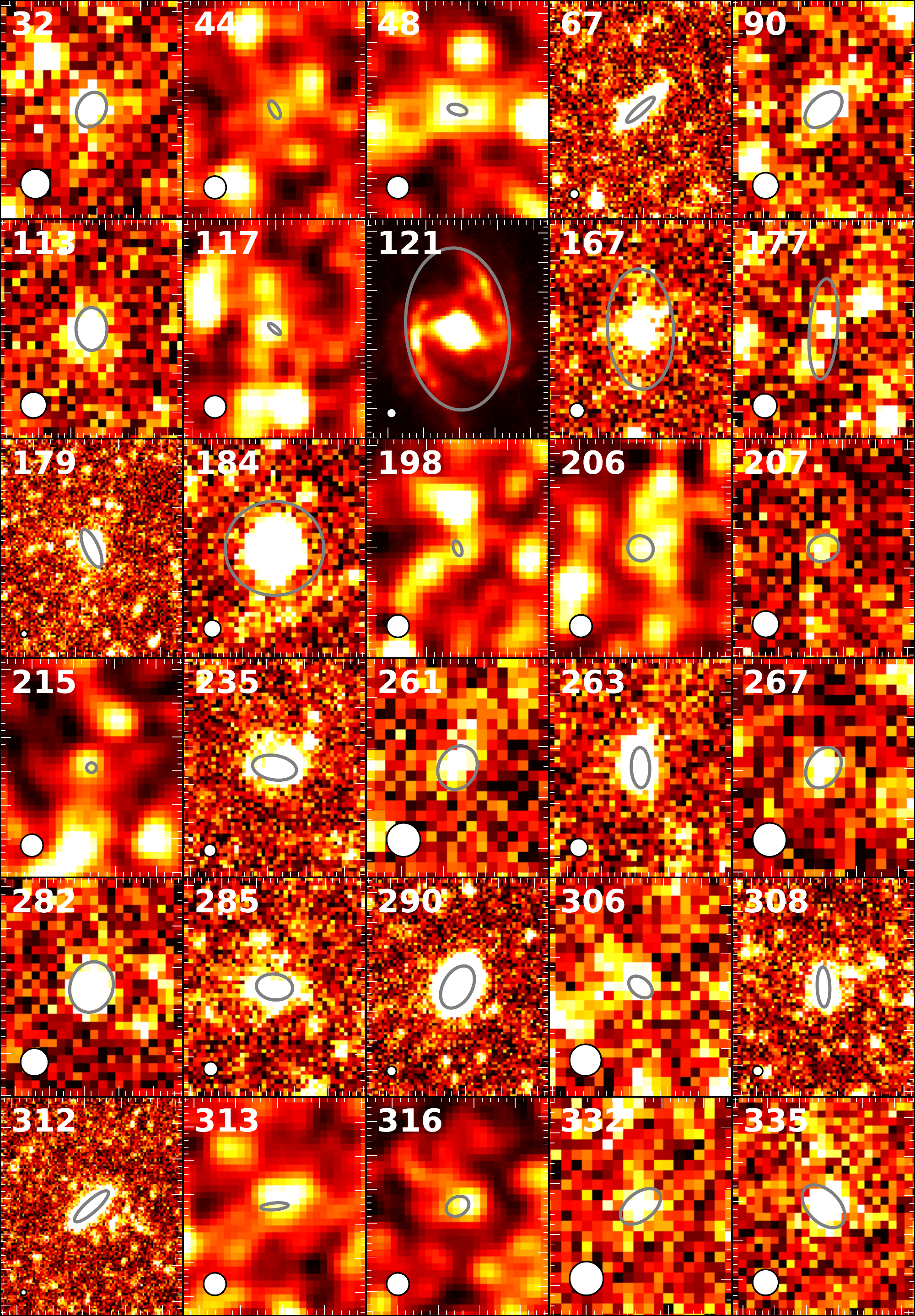}
\caption{The 30 HeFoCS galaxies detected at 250\,$\micron$. The beam size is shown in the lower left hand corner. The grey ellipses indicate the optical extent ($D_{25}$) of each galaxy.}
\label{fig:detected250}
\end{figure*}

In this section we describe the HeFoCS detection rate in each $Herschel$ band and compare our results to those obtained for galaxies in Virgo by~\citep{auld12}. We then investigate the location of FIR detected and undetected galaxies within the cluster. Every HeFoCS galaxy detected at 250\,$\micron$ is shown in Figure~\ref{fig:detected250}, where the grey ellipse shows the extent and location of the optical counterpart. 

\subsection{Detection rates}
\label{sec:detection_rates}
\begin{table}
\centering
  \begin{tabular}{ccc}
\hline
  Band $\micron$ & Number of detections (N) & Detection rate (\%) \\ \hline
  100      & 19            & 9          \\ 
  160      & 18            & 8          \\
  250      & 30            & 13          \\
  350      & 28            & 12          \\
  500      & 21            & 9          \\
\hline
  \end{tabular}
  \caption {Detection rates of all the FCC galaxies in the $Herschel$ bands. 237 FCC galaxies fall into the SPIRE maps and 200 fall into both PACS and SPIRE in total.}
\label{tab:det}
\end{table}

\begin{table*}
\centering
\begin{tabular}{ccccccc} 
\hline
Morphological  &  \multicolumn{3}{c}{Virgo} & \multicolumn{3}{c}{Fornax} \\ \hline
Type & Total & Detected & \% & Total & Detected & \%\\
dE/dS0 & 314 & 14 & 4\,$\pm$\,1 & 185 & 11 & 6\,$\pm$\,2 \\
E/S0 & 86 & 29 & 34\,$\pm$\,6 & 29 & 6 & 21\,$\pm$\,8 \\
Sa/Sb/Sc/Sd & 152 & 138 & 91\,$\pm$\,8 & 10 & 9 & 90\,$\pm$\,30 \\
BCD/Sm/Im/dS & 157 & 74 & 47\,$\pm$\,5 & 13 & 4 & 31\,$\pm$\,15 \\

\hline 
\end{tabular} 
  \caption{A comparison of detection rates in the SPIRE 250\,$\micron$ band, between the Virgo and Fornax clusters. Errors are simply root N. The galaxies have been split into dwarf (dE\,/\,dS0), early (E\,/\,S0), late (Sa\,/\,Sb/\,Sc/\,Sd), and irregular (BCD\,/\,Sm\,/\,Im\,/dS).}
  \label{tab:morh}
\end{table*} 

\begin{figure}
\centering
\includegraphics[width=\linewidth]{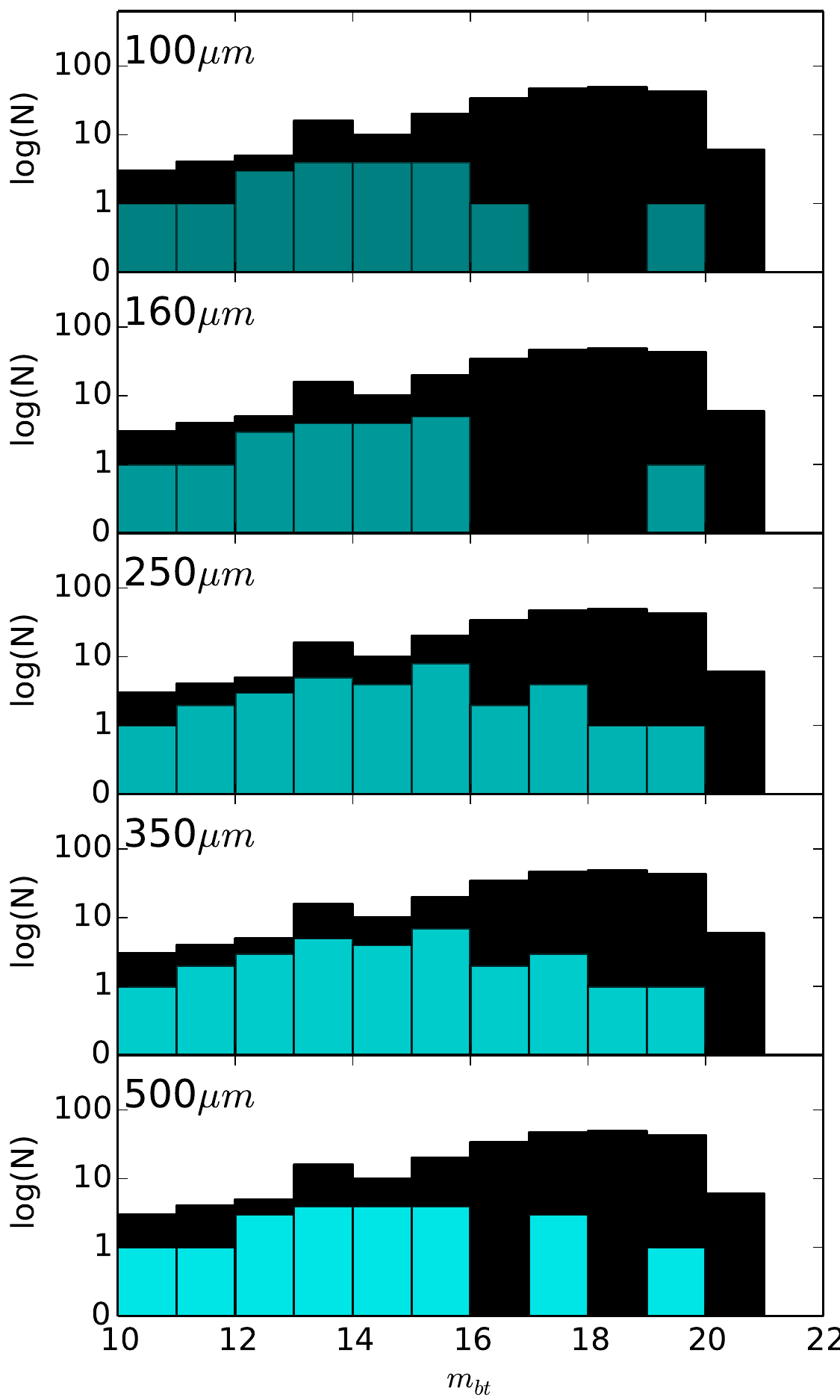}
\caption{A histogram of optical magnitude $m_{bt}$ of the FCC galaxies - the black and cyan bars are the total and FIR detected galaxies respectively.}
\label{fig:dopt}
\end{figure}
\begin{figure}
\centering
\includegraphics[trim= 0 0 0 0, width=\linewidth]{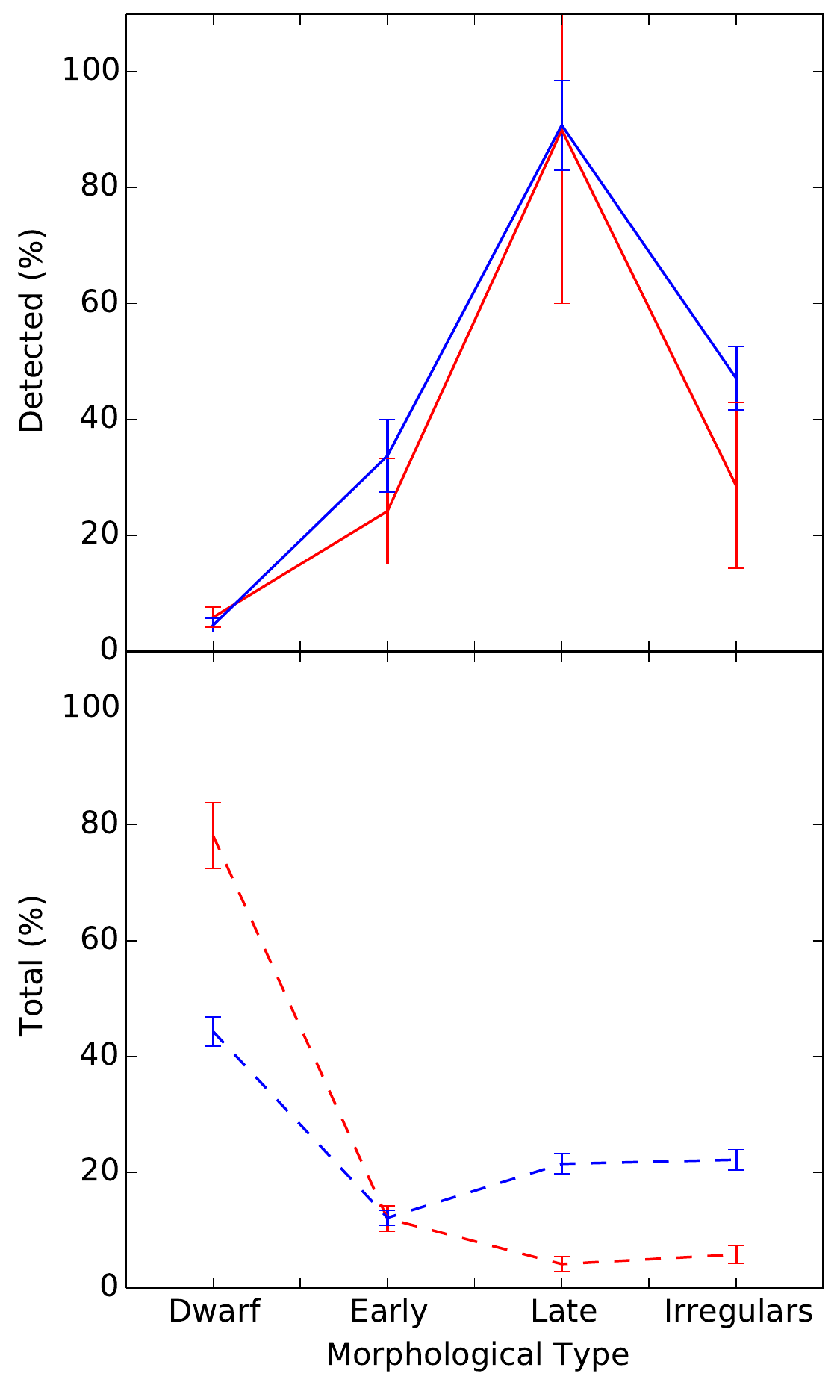}
\caption{The morphologies of the Fornax and Virgo galaxies in red and blue, respectively. The upper panel represents the percentage detected in SPIRE 250\,$\micron$ band. The lower panel shows the overall morphological make up of each cluster. The bins are as follows; dwarf (dE\,/\,dS0), early (E\,/\,S0), late (Sa\,/\,Sb/\,Sc/\,Sd), and irregulars (BCD\,/\,Sm\,/\,Im\,/dS).}
\label{fig:morph}
\end{figure}

Figure~\ref{fig:dopt} shows the distribution of optical magnitudes $m_{bt}$ of all (black) and detected (blue) FCC galaxies. Except for one faint galaxy (discussed separately in Section 4.1.1), no galaxies are detected in the FIR below $m_{bt}$ = 18.2. Therefore, we do not expect that a deeper optical catalogue would increase the number of FIR detections in our current data.

Table~\ref{tab:det} indicates how many galaxies were recovered in each band above a 3$\sigma$ noise level in the FIR maps. The SPIRE bands have higher detection rates than the PACS bands, and 250\,$\micron$ has the highest detection rate of all. This is due to a combination of its sensitivity and the typical shape of the FIR SED. Consequently, we use the 250\,$\micron$ band to compare Fornax and Virgo. At 250\,$\micron$ we detect 30 of 237 (13\,\%) FCC galaxies. This is significantly less than in Virgo, where 254 of 750 (34\,\%) VCC galaxies are detected~\citet{auld12}.

In order to investigate the source of the lower global detection rates in Fornax in comparison to Virgo, we examine the morphological make up of each cluster and the detection rates therein. We separate the galaxies into 1 of 4 morphological groups; dwarf (dE\,/dS0), early (E\,/\,S0), late (Sa\,/\,Sb/\,Sc\,/\,Sd), and irregular (BCD\,/\,Sm\,/\,Im\,/\,dS). The upper panel in Figure~\ref{fig:morph} shows the fraction of galaxies detected in each morphological group, while the lower panel shows the overall morphological make up of each cluster (tabulated in Table~\ref{tab:morh}). Dwarf galaxies are the most numerous in both clusters, however, only 4\,\% and 6\,\% are recovered at 250\,$\micron$ for Virgo and Fornax, respectively. Early, late\footnote{FCC\,176 was originally classified as Sa by~\citet{ferguson90}, however, we did not detect this galaxy in any $Herschel$ bands. Upon further inspection it has a very red colour, B-V = 0.88~\citep{prugniel98}, placing it well within the red sequence, it has also been reclassified S0, more latterly by~\citet{deVaucouleurs91}, we have adopted this reclassification.}, and irregular-type galaxies are detected at 21\,\%, 90\,\%, and 31\,\% in Fornax and 34\,\%, 91\,\%, and 47\,\% in Virgo, respectively. The lower panel of Figure~\ref{fig:morph} shows that Fornax has a far higher fraction of dwarf galaxies, with the lowest detection rate, and far less late and irregular-type type galaxies with the highest detection rate. Furthermore Figure~\ref{fig:morph} shows the fraction of early-type galaxies is the same in both clusters, having no effect on the global detection rate.  What is remarkable, is that within the errors the two clusters match each other very closely with respect to the fraction of detected galaxies in each morphological group. The above implies that the lower global detection rates in Fornax are tracing the morphological make up of the cluster.

\begin{figure*}
\centering
\includegraphics[trim = 0 0  0 0, width=162mm]{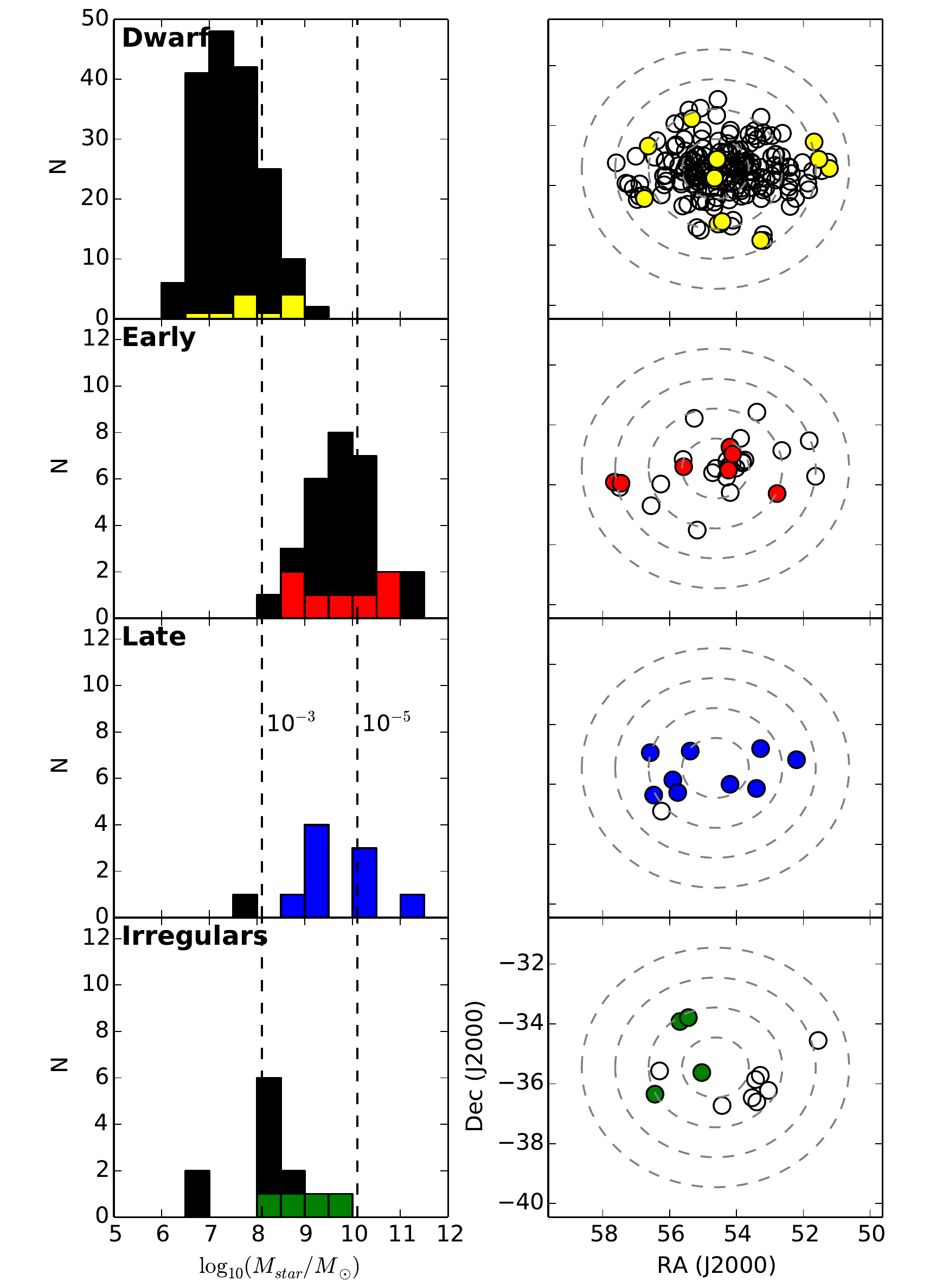}
\caption{Histograms of stellar mass for 4 morphological types; dwarf (dE\,/dS0), early (E\,/\,S0), late (Sa\,/\,Sb/\,Sc\,/\,Sd), and irregular (BCD\,/\,Sm\,/\,Im\,/dS). The black and coloured histograms are for undetected and detected galaxies at 250\,$\micron$, respectively. The vertical dashed lines represent our estimated stellar mass detection limits for the indicated dust-to-stars mass ratio. Note the change in the Y-scale for the dwarf galaxies panel. The adjacent plots show the locations within the cluster of the undetected and detected galaxies, with empty and filled markers respectively.}
\label{fig:dsmass}
\end{figure*}

In order to better understand the limits of our data we estimate the limiting dust mass required for a detection at 250\,$\micron$. The lowest detected 250\,$\micron$ flux in our FIR catalogue is $\sim$\,15\,mJy. Using the relation calibrated in Section~\ref{sec:250umtodust}, the corresponding limiting dust mass is log($M_{Dust}$/M$_{\odot}$)\,=\,5.1. If we assume that early and late-type galaxies typically have dust-to-stellar mass ratios of approximately, $\log(M_{Dust}/M_{Stars}) = -5$ and -3, respectively~\citep{cortese12,smith12}, then we should detect galaxies with stellar masses of log($M_{Stars}$/M$_{\odot}) \ge$ 10.1 and 8.1, respectively. We can see this more clearly in Figure~\ref{fig:dsmass}. The 4 left-hand panels display the distribution of stellar mass in the 4 morphological groups described above, with black and coloured histograms showing galaxies undetected and detected at 250\,$\micron$, respectively. The dashed lines indicate the stellar mass above which we expect to detect galaxies with a dust-to-stellar mass ratio of $\log(M_{Dust}/M_{Stars}) = -5$ and $-3$, respectively.

The dwarf galaxies are the most challenging morphological group to detect in the FIR, due to their low stellar masses, thus, requiring a substantially higher dust-to-stellar mass ratio for their detection. 6\%(11) of dwarf galaxies are detected, whereas we would expect to detect 18\%(33) of the dwarf galaxies if they had dust-to-stellar mass ratios of $\log(M_{Dust}/M_{Stars}) = -3$ similar to a typical late-type galaxy. The righthand panel of Figure~\ref{fig:dsmass}, shows where these galaxies are projected spatially in the cluster. The FIR detected dwarf galaxies generally appear on the outskirts of the cluster. To quantify this, Table~\ref{tab:morh-loc} lists the average projected cluster centric radius as a fraction of the virial radius ($R_{virial}$ =0.7\,Mpc) for FIR detected and undetected galaxies. On average detected dwarf galaxies are found at a 0.84\,R$_{virial}$, whereas undetected are at 0.50\,R$_{virial}$. These detected dwarf galaxies are found on the outskirts of the cluster in a similar position to the transition dwarfs identified in the Virgo cluster by~\citet{delooze13}. 

Only 21\,\% of all early-type galaxies are detected by $Herschel$ at 250\,$\micron$, with some of the extremely dust deficient early types having dust-to-stars ratios of below $\log(M_{Dust}/M_{Stars}) = -6.6$. Early-type galaxies appear very centrally concentrated when compared to the dwarf and irregular-type galaxies. However, both detected and undetected galaxies are found at an average projected cluster centric radius of $\sim 0.5$R$_{virial}$. It would appear that cluster centric radius has no perceivable effect on whether or not an early-type galaxy is detected by $Herschel$. 

There are nine late-type galaxies in the Fornax cluster and they are all detected except for FCC\,299. In order to be detected at 250\,$\micron$, the latter would require a dust-to-stars ratio greater than $\log(M_{Dust}/M_{Stars}) = -3$ due to its low stellar mass of $\log(M_{Stars}$/M$_{\odot}$) = 7.8. The detected galaxies have a mean projected cluster centric radius of $\sim 0.56$\,R$_{virial}$ and no late-type galaxy has a radius less than 0.3\,R$_{virial}$.

The majority of irregular-type galaxies would be detected at 250\,$\micron$ if they had dust-to-stars ratios of $\log(M_{Dust}/M_{Stars}) = -3$. Instead, approximately 31\,\% of the irregular-type galaxies are detected, preferentially with higher stellar masses. There is no obvious trend to where they are located in the cluster, both detected and undetected galaxies having a mean projected radius of $\sim 0.6$\,R$_{virial}$.

\begin{table}
\centering
\begin{tabular}{ccc} 
\hline
Morphological & $<R_{detected}> $ & $<R_{undetected}>$ \\ 
Type &($R/ R_{virial}$)&($R/ R_{virial}$)\\ \hline
Dwarf&0.84\,$\pm$\,0.04&0.50\,$\pm$\,0.01\\
Early&0.56\,$\pm$\,0.05&0.50\,$\pm$\,0.02\\
Late&0.56\,$\pm$\,0.02&0.83\\
Irregulars&0.61\,$\pm$\,0.06&0.57\,$\pm$\,0.01\\
\hline
\end{tabular} 
  \caption{A comparison of galaxies detected and undetected in the SPIRE 250\,$\micron$ band. Projected radii are given as a fraction of the Fornax cluster virial radius of 0.7\,Mpc~\citep{drinkwater01}.}
  \label{tab:morh-loc}
\end{table} 

\subsubsection{FCC\,215}

From 185 dwarf galaxies identified in the FCC, only 11 were detected in the 250\,$\micron$ band, and only FCC\,215 was detected in 3 or more $Herschel$ bands. FCC\,215 has a very high dust-to-stars ratio (approximately $\log(M_{Dust}/M_{Stars}) = -1$) and a very faint optical magnitude ($m_{bt} \sim 19$), making it an interesting object worthy of further inspection. 

FCC\,215 has a dust mass of $\log_{10}(M_{dust}/M_{\odot}) = 5.2$ and a stellar mass of $\log_{10}(M_{stars}/M_{\odot} )= 6.5$. It is just detected in the 3 SPIRE bands at S/N $\le$ 5. The SED fit is quite poor with $\chi^{2}_{dof=3} = 9.94$. The SED appears very flat, which may indicate it is a background galaxy with a synchrotron component. However, it is listed in NED as having a velocity of 1964\,km\,s$^{-1}$, which places it inside the cluster. Its optical colour is very blue, B-R = 0.36, suggesting that the galaxy is undergoing/has undergone an episode of recent star formation. Assuming that this is a bonafide detection, how could it have such a high dust-to-stars ratio? Is it possible for a galaxy to produce this much dust? Using a closed box model of a galaxy, i.e. no inflow or outflow of material,~\citet{edmunds98} derive; $\Delta_{max, f} = \eta p f \log(1/f)$, where $\Delta_{max, f}$ is the maximum mass of dust a galaxy could possess with a gas fraction $f$, a fraction of metals in the dust $\eta$ and a stellar yield $p$. The stellar yield is the fraction of metals produced per unit mass of gas freshly formed in nucleosynthesis. Its value has been estimated to lie between 0.004 and 0.0012~\citep{vila92}. The fraction of metals in the dust $\eta$ has been estimated by~\cite{meyer98} and more latterly by~\citet{Davies14} as 0.5. The gas fraction is $f = M_{gas} / (M_{stars} + M_{dust} + M_{gas})$, so using the equation above we can estimate the gas mass required, for FCC\,215 to have a dust-to-stars ratio of $\log(M_{Dust}/M_{Stars}) = -1.5$. The gas-to-stars ratio would have to be 1 and thus a gas mass of $\log_{10}(M_{gas}$/M$_{\odot}) = 6.5$, making it also very gas rich. Currently the only 21cm survey that covers this region of sky is the \hi\ Parks All Sky Survey (HIPASS)~\citep{hipass}. HIPASS does not detect FCC\,215, yet their estimated rms noise of $\sim$15\,mJy\,beam$^{-1}$ approximately corresponds to an \hi\ gas mass $\log_{10}(M_{HI}$/M$_{\odot}) = 8$ at the distance of Fornax, meaning that HIPASS would be unable to detect FCC\,215 even if all the gas content was locked up in \hi. The HeFoCS has secured time to map the Fornax cluster, using the Australia Telescope Compact Array. The estimated survey detection limit is M$_{HI} \simeq 10^{7}$ M$_{\odot}$ at the distance of Fornax, very close to our predicted upper estimate of the gas mass of FCC\,215. 
 
\subsection{Analysis of SED fits, dust masses \& tempertures}
\label{sec:sed}

\subsubsection{Environmental effect on dust in galaxies}

\begin{table*}
\centering
\begin{tabular}{cccccc}
\hline
\\
Sample 1 & Sample 2 & $\mu_{1}$($\sigma_{1}$) & $\mu_{2}$($\sigma_{2}$) &\multicolumn{2}{c}{K-S test} \\
~&~&~&~&Value& $P_{value}$
\\ \hline \\
\multicolumn{6}{l}{\textbf{Dust Mass (log($M_{Dust}$/$M_{\odot}$))}} \\
Virgo Early&Virgo Late&6.18(0.12)&6.68(0.06)&0.488&0.011\\
Fornax Early&Fornax Late&5.82(0.2)&6.54(0.19)&0.571&0.113\\
Virgo Early&Fornax Early&6.18(0.12)&5.82(0.2)&0.6&0.102\\
Virgo Late&Fornax Late&6.68(0.06)&6.54(0.19)&0.179&0.784\\
\\
\multicolumn{6}{l}{\textbf{ Dust Mass / Stellar Mass ($M_{Dust}$ / $M_{Stellar}$)}} \\
Virgo Early&Virgo Late&-3.62(0.19)&-2.76(0.04)&0.611&0.001\\
Fornax Early&Fornax Late&-3.89(0.27)&-2.94(0.08)&0.929&0.001\\
Virgo Early&Fornax Early&-3.62(0.19)&-3.89(0.27)&0.364&0.645\\
Virgo Late&Fornax Late&-2.76(0.04)&-2.94(0.08)&0.332&0.104\\
\\
\multicolumn{6}{l}{\textbf{Dust Temp. (K)}} \\
Virgo Early&Virgo Late&21.65(0.94)&19.27(0.24)&0.442&0.028\\
Fornax Early&Fornax Late&20.82(1.77)&17.47(0.96)&0.443&0.355\\
Virgo Early&Fornax Early&21.65(0.94)&20.82(1.77)&0.236&0.975\\
Virgo Late&Fornax Late&19.27(0.24)&17.47(0.96)&0.4&0.027\\
\hline
\end{tabular}
\caption{A statistical comparison of Fornax and Virgo galaxies using dust mass, dust-to-stellar mass, and dust temperature. Early types include E and S0, while all other galaxy types are classified as `late'.}
\label{tab:stats}
\end{table*}

For the following analysis we split the sample into early and late-type galaxies and initially consider only the 22 galaxies detected in at least 3 $Herschel$ bands. `Early' was classified as anything earlier than Sa and `late' as anything later than (and including) Sa. The SED of all galaxies was fitted with a single temperature modified blackbody with $\beta=2$. Only two galaxies, FCC\,215 (discussed above) and FCC\,306, were poorly fitted using this emissivity, with $\chi^{2}_{dof = 3} =$ 9.94 and 18.65. The average for the entire sample was, $<\chi^{2}_{dof = 3}> = 2.92$. If FCC\,215 and 306 are removed, then the average for the sample falls to $<\chi^{2}_{dof = 3}> = 1.78$. In Table~\ref{tab:FIRmasses} we include all galaxies with measured dust mass and temperature. Figure~\ref{fig:sed_fits} shows the SED fits for each galaxy.

Detected late-type galaxies have dust masses ranging from $\log_{10}(M_{dust}$/M$_{\odot}) = 5.5$ to $8.2$ and temperatures of 11.2 to 23.7\,K, with mean values $\log_{10}(M_{dust}$/M$_{\odot}) = 6.5$ and 17.5\,K. By contrast, detected early types have a narrower range of dust masses of $\log_{10}(M_{dust}$/M$_{\odot}) = 5.4$ to $6.6$ and temperatures of 14.9 to 25.8\,K, with mean values $\log_{10}(M_{dust}$/M$_{\odot}) = 5.8$ and 19.3\,K. Detected Fornax galaxies have mean dust-to-stellar mass ratios of $\log_{10}(M_{dust}/M_{stars})$ = -3.87 and -2.93, for early and late-types, respectively. As expected from our previous results for Virgo, late-types have a richer and cooler dust reservoir, and early-types have a relatively depleted and warmer ISM.

In Table~\ref{tab:stats} we use the Kolmogorov-Smirnov two sample test (KS) to make a more quantitative comparison between Virgo and Fornax's early and late-type galaxy populations with respect to dust mass, dust-to-stars ratio, and dust temperature. Here we use 140 of the~\citet{auld12} galaxies that had SEDs modelled identically to our sample using a single temperature component with a fixed $\beta = 2$ emissivity. For Virgo we use stellar masses calculated using H band magnitudes and SDSS g-r colours from~\citet{Davies14}. Virgo, like Fornax, has early types that have lower dust masses and higher temperatures than its late types. However, a KS test shows that for a given morphological type the FIR properties of galaxies in Fornax and Virgo are statistically identical (with the caveat that we are only sampling the massive galaxies, $\log_{10}(M_{star}$/M$_{\odot}) \ge 8.2$). The above results suggest that the different cluster environments have had very little effect on the dust properties of early or late-type galaxies.

~\citet{auld12} compared the Virgo cluster to the Herschel Reference Survey~\citep[HRS;][]{Boselli10,hrslate,smith12}. The HRS is a volume limited (15\,$\le$ D $\le$\,25), K band (K $\ge 8.7$) selected sample. It covers a range of environments from the field to the core of the Virgo cluster, making it an ideal comparison sample.~\citet{auld12} showed that early-type galaxies in the Virgo cluster and HRS field show very similar dust properties. However, late-type galaxies typically have larger dust masses in the field.~\citet{auld12} concluded that the difference in dust mass between field and cluster late-type galaxies was due to dust removal in the cluster environment. The implication of this result as well as the results presented in this paper, is that early-type galaxies appear identical in their FIR properties irrespective of what environment they originated from. Furthermore, the larger dust reservoirs of late-type galaxies in the field and the lack of difference in FIR properties between Fornax and Virgo, suggests that this change in dust mass likely occurred before they entered the cluster environment. 

It is worth noting that `global' environment on its own may not be the best tracer of the action of physical processes. A quantity more sensitive to direct interaction with the cluster environment is the \hi-deficiency.~\citet{hrslate} compare the FIR properties of galaxies, separated in both \hi-deficiency and global environment. They found an $\sim$\,$8 \sigma$ difference in $\log(M_{Dust}/M_{Stars}$) when comparing \hi-normal and \hi-deficient galaxies, whereas only a $\sim$\,$3 \sigma$ difference is found between samples separated based on the environment (i.e. field and cluster members)

\subsubsection{Orgin of dust in galaxies}

\begin{figure}
\centering
\includegraphics[width=\linewidth]{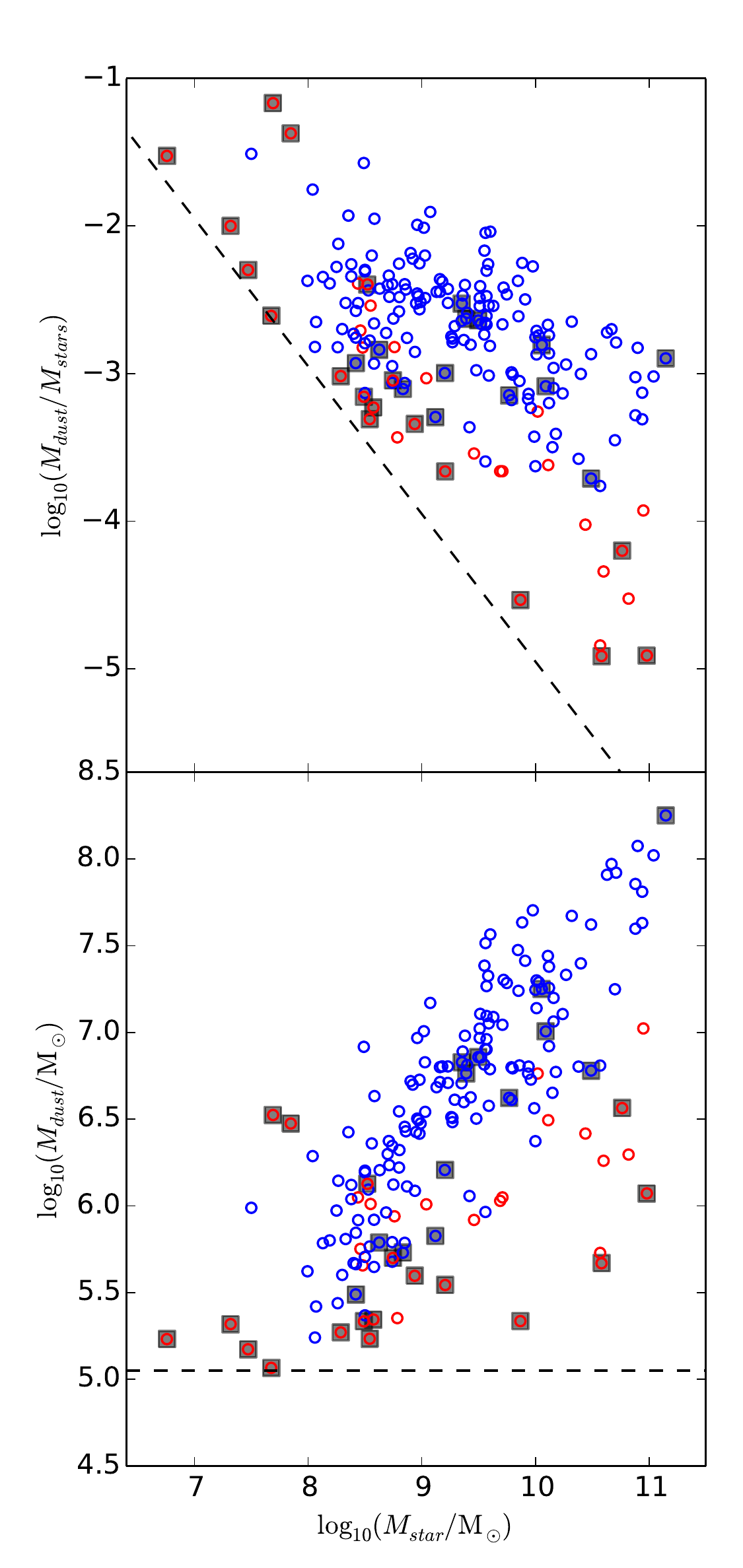}
\caption{The upper pannel shows stellar mass against dust-to-stars ratio, the lower pannel shows stellar against dust mass, for both Virgo and Fornax, where the latter are designated by a marker set onto a black square. Early and late-type galaxies are shown as red and blue markers, respectively. The dashed line represents the minimum dust mass and dust-to-stellar mass we can detect for a given stellar mass, for the lower and upper pannels, respectively.}
\label{fig:stellardustmass}
\end{figure}

In order to extend our analysis of dust and stellar mass to lower limits, and to study how the dust-to-stars ratio changes with lower dust and stellar masses (Figure~\ref{fig:stellardustmass}), an additional 9 galaxies were included in the analysis. These galaxies had insufficient SED data to be fitted by a modified blackbody and so the 250$\micron$ flux density was used as a proxy for dust mass (described in Section~\ref{sec:250umtodust}). The diagonal dash line in Figure~\ref{fig:stellardustmass} indicates the minimum dust-to-stellar mass detected, given our previous estimate of a minimum detectable dust mass of $\log_{10}(M_{dust}$/M$_{\odot})$ = 5.1 (Section~\ref{sec:detection_rates}). The same morphological categories are used - `early' was classified as anything earlier than Sa and `late' as anything later than, and including, Sa.

Figure~\ref{fig:stellardustmass} shows early and late-type galaxies designated by red and blue markers, respectively, from both clusters. Fornax galaxies are indicated by a marker set onto a black square. We have measured the correlation between $M_{dust}$ and $M_{star}$ using the Pearson correlation coefficient (PCC). Late-type galaxies have a PCC of 0.84, early-type galaxies have a PCC of 0.43. The correlation between dust and stellar mass in late-type galaxies most likely finds its origins in the mass-metallicity relation~\citep{lequeux79, tremonti04, lara10,hughes13}. These authors have shown that gas phase metallicity correlates with stellar mass and so we might also expect this to be true for the metals in the dust.

Early-type galaxies in Fornax and Virgo have a very large range of dust-to-stellar mass ratios, $-1.3 \ge \log_{10}(M_{star}/M_{dust}) \ge -6.2$, and the weak correlation of dust to stellar mass could be due to the imposed limiting dust mass, artificially creating a correlation, as shown in Figure~\ref{fig:stellardustmass}. However, the PCC for early-type galaxies is far lower than we found for late-type galaxies, implying, that stellar mass is far less if at all correlated with dust mass in early-type galaxies.

A clue about origin of these two different correlations may lie in the distribution of the dust within early and late-type galaxies. We have calculated the ratio of the FIR to optical size for Fornax cluster galaxies, where this ratio is defined as the FIR diameter of emission $D_{FIR}$ as defined in Section~\ref{sec:sourcemeasurement} divided by the optical diameter $D_{25}$. We will use the FIR diameter as measured at 250\,$\micron$, and thus we will for the rest of this section refer to $D_{FIR}$ as $D_{250}$. Only 1 of 7 early and 2 of 13 late types have FIR emission that is smaller than the FWHM of the $Herschel$ 250\,$\micron$ beam, and are thus measured as point sources with a $D_{250}$ equal to the 250\,$\micron$ beam size. As the majority have $D_{250}$ greater than the PSF FWHM, we can use them to measure the distribution of the dust in comparison to the stars. The FIR/optical size ratio is 0.464 and 0.903, for early and late-type galaxies, respectively. In order to further test the effect of the $Herschel$ beam we restricted the sample to galaxies with an optical diameter greater than 3 times the 250\,$\micron$ beam FWHM. This results in a FIR/optical size ratio of 0.305 and 0.917 for early and late-type galaxies, respectively, thus showing that the beam size has a limited effect on the overall result. This shows that dust in early-type galaxies is very centrally concentrated in comparison to late-type galaxies. This has been demonstrated previously for early-type galaxies by~\citet{smith12} and~\citet{alighieri13}.~\citet{cortese10} showed that \hi-deficient galaxies in the cluster environment also had smaller FIR/optical size ratios, suggesting that dust had been stripped from these galaxies, this would affect the late-type galaxies far more than early-type galaxies, as the above shows that dust in late-type galaxies is held less deeply in their potential wells.

~\citet{davis13} and~\citet{helfer03} measure the extent of molecular hydrogen in early-type and late-type galaxies. They find the size ratio of molecular gas to optical radius as $\sim 0.25$ in both cases. In the case of late-types, the spatial extent of molecular gas is much less than of the dust, whereas early-type galaxies have molecular gas and dust that appear spatially coincident. This suggests that the origin of dust in early-types may be the same as that of molecular gas (see below). Whereas dust in late-types is coincident with the stellar population, and as shown above, dust mass is regulated by stellar mass (i.e. mass-metallicity relation), suggesting an internal origin. 

Dust in early-type galaxies has two possible origins, either internally produced in the atmospheres of evolved stars~\citep{whittet92} and supernovae remnants~\citep{morgan03}, or externally obtained from mergers with other galaxies. The strongest prediction for dust created internally is that the mass of dust and stars should be spatially correlated. Figure~\ref{fig:stellardustmass} shows that this is clearly not the case for early-type galaxies. The stellar population must produce dust, but~\citet{clemens10} show that it is destroyed on a short timescale of \textless 50\,Myrs. They argue that this is far shorter than the dust-transfer timescale, and thus dust created in outer regions of a galaxy is effectively destroyed ``on-the-spot''. However, the dust destruction timescale can be greatly extended if the dust is embedded in a cloud of molecular hydrogen, leading to lifetimes of a few 100\,Myrs~\citep{jones11}. This indicates that dust created internally cannot be the main source of dust in early-type galaxies.

If the dominant source of dust is not internal,~\citet{smith12} argue that it may have an external origin such as mergers with dust rich galaxies. Mergers of different dust masses at different times would explain the large range of dust-to-stars ratios seen in early-type galaxies as well as the $\sim 75 \%$ of systems which we do not detect with $Herschel$. However, as shown above, the FIR properties of early-type galaxies do not change between Virgo and Fornax (Table~\ref{tab:stats}) or the HRS field~\citep{auld12}, suggesting that the flow of dust into and out of these systems must be invariant with environment. Since~\citet{clemens10} show that the destruction time in early-type galaxies is determined by thermal sputtering, and thus is largely independent of the environment, our findings would imply that the merger rate is roughly the same in all three environments. This is at odds with the idea that the merger rate depends on environment. For example, \citet{mihos04} shows that mergers are far less common in clusters than in groups or in the field - thus there is a dilemma.

The mystery deepens if we compare our FIR results for early-types to the molecular gas component  of the ISM.~\citet{davis11} show that the detection rate of the molecular ISM and the molecular gas-to-stars ratios for early-type galaxies are invariant to environment, mirroring the FIR results presented in this paper. However, they discovered that the gas kinematics inside and outside of clusters is different. They found one third of galaxies outside of clusters had gas kinematically misaligned to their stars, supporting an external origin. Interestingly, this was not see in early-types inside the cluster. 

\section{Summary}

We have undertaken the deepest FIR survey of the Fornax  cluster using the $Herschel$ Space Observatory. Our survey covers over 16 deg$^2$ in 5 bands and extends to the virial radius of the cluster, including 237 of the 340 FCC galaxies. We have used the optical positions and parameters of these FCC galaxies to fit appropriate apertures to measure FIR emission. We have detected 30 of 237 (13\,\%) cluster galaxies in the SPIRE 250\,$\micron$ band, a significantly lower detection rate than in the Virgo cluster~\cite[34\,\%; see][]{auld12}. 

In order to better understand the global detection rate we separated Fornax and Virgo galaxies into 4 morphological categories: dwarf (dE\,/dS0), early (E\,/\,S0), late (Sa\,/\,Sb/\,Sc\,/\,Sd), and irregular (BCD\,/\,Sm\,/\,Im\,/\,dS). We examined the detection rate for each morphological group in the 250\,$\micron$ band as it has the highest detection rate of all the $Herschel$ bands. In Fornax we detect 6\%, 21\%, 90\%, and 31\% of dwarf, early, late, and irregular, respectively. These results agrees with the fraction of detected galaxies in each morphological category in the Virgo cluster, indicating that the lower global detection rate in Fornax is due to its lower fraction of late-type galaxies. 

For galaxies detected in at least 3 bands we fit a modified blackbody with a fixed beta emissivity index of 2, giving dust masses and temperatures for 22 Fornax galaxies. Fornax's early-type galaxies show lower dust masses and hotter temperatures than late-type galaxies. When comparing early-type galaxies from the Fornax cluster to their counter-parts in the Virgo cluster, their FIR properties are statistically identical. The same is true for the late-type galaxies. This may suggest that the effect of the cluster is more subtle than previously thought and that the evolution of the ISM components has mostly taken place before the cluster was assembled.

We observe dust mass to be well correlated to stellar mass for late-type galaxies. We suggest that this correlation has its origins in the mass-metallicity relation~\citep{lequeux79, tremonti04,lara10,hughes13}, as the ratio between the mass of metals in the dust and the gas has been found to be 0.5~\citep{meyer98,Davies14}. It therefore follows that any correlation with gas phase metallicity should also be observed between stellar and dust mass.

We find early-type galaxies to have a very large range of dust-to-stars ratios, $-1.3 \ge \log_{10}(M_{star}/M_{dust}) \ge -6.2$. We argue that this supports a scenario where the dust in early-type galaxies is from an external origin, as has been previously suggested by other authors~\citep{smith12}. As FIR properties are statistically identical between environments, therefore so must the balance between dust input/creation and removal/destruction. However, this conclusion is perplexing as mergers are thought to be far less common in clusters when compared to groups or the field~\citep{mihos04}, and dust destruction is largely regulated internally~\citep{clemens10}, thus invariant with respect to environment.

\section*{Acknowledgements}
SPIRE has been developed by a consortium of institutes led by
Cardiff University (UK) and including University of Lethbridge
(Canada); NAOC (China); CEA, LAM (France); IFSI, University
of Padua (Italy); IAC (Spain); Stockholm Observatory (Sweden);
Imperial College London, RAL, UCL-MSSL, UKATC, University
of Sussex (UK); and Caltech, JPL, NHSC, University of Colorado
(USA). This development has been supported by national funding
agencies: CSA (Canada); NAOC (China); CEA, CNES, CNRS
(France); ASI (Italy); MCINN (Spain); SNSB (Sweden); STFC
(UK); and NASA (USA). HIPE is a joint development (are joint
developments) by the Herschel Science Ground Segment Consortium,
consisting of ESA, the NASA Herschel Science Center, and
the HIFI, PACS and SPIRE consortia.
The research leading to these results has received funding
from the European Community’s Seventh Framework Programme
(/FP7/2007–2013/) under grant agreement No. 229517.
This research has made use of data obtained from the SuperCOSMOS Science Archive, prepared and hosted by the Wide Field Astronomy Unit, Institute for Astronomy, University of Edinburgh, which is funded by the UK Science and Technology Facilities Council.
We acknowledge the usage of the HyperLeda database (http://leda.univ-lyon1.fr). IDL is a postdoctoral researcher of the FWO-Vlaanderen (Belgium). We gratefully acknowledge the contribution of L. Cortese to this work.
The Parkes telescope is part of the Australia Telescope which is funded by the Commonwealth of Australia for operation as a National Facility managed by CSIRO.
\def\ref@jnl#1{{\rmfamily #1}}% 
\newcommand\aj{\ref@jnl{AJ}}% 
          % Astronomical Journal 
\newcommand\araa{\ref@jnl{ARA\&A}}% 
          % Annual Review of Astron and Astrophys 
\newcommand\apj{\ref@jnl{ApJ}}% 
          % Astrophysical Journal 
\newcommand\apjl{\ref@jnl{ApJ}}% 
          % Astrophysical Journal, Letters 
\newcommand\apjs{\ref@jnl{ApJS}}% 
          % Astrophysical Journal, Supplement 
\newcommand\ao{\ref@jnl{Appl.~Opt.}}% 
          % Applied Optics 
\newcommand\apss{\ref@jnl{Ap\&SS}}% 
          % Astrophysics and Space Science 
\newcommand\aap{\ref@jnl{A\&A}}% 
          % Astronomy and Astrophysics 
\newcommand\aapr{\ref@jnl{A\&A~Rev.}}% 
          % Astronomy and Astrophysics Reviews 
\newcommand\aaps{\ref@jnl{A\&AS}}% 
          % Astronomy and Astrophysics, Supplement 
\newcommand\azh{\ref@jnl{AZh}}% 
          % Astronomicheskii Zhurnal 
\newcommand\baas{\ref@jnl{BAAS}}% 
          % Bulletin of the AAS 
\newcommand\jrasc{\ref@jnl{JRASC}}% 
          % Journal of the RAS of Canada 
\newcommand\memras{\ref@jnl{MmRAS}}% 
          % Memoirs of the RAS 
\newcommand\mnras{\ref@jnl{MNRAS}}% 
          % Monthly Notices of the RAS 
\newcommand\pra{\ref@jnl{Phys.~Rev.~A}}% 
          % Physical Review A: General Physics 
\newcommand\prb{\ref@jnl{Phys.~Rev.~B}}% 
          % Physical Review B: Solid State 
\newcommand\prc{\ref@jnl{Phys.~Rev.~C}}% 
          % Physical Review C 
\newcommand\prd{\ref@jnl{Phys.~Rev.~D}}% 
          % Physical Review D 
\newcommand\pre{\ref@jnl{Phys.~Rev.~E}}% 
          % Physical Review E 
\newcommand\prl{\ref@jnl{Phys.~Rev.~Lett.}}% 
          % Physical Review Letters 
\newcommand\pasp{\ref@jnl{PASP}}% 
          % Publications of the ASP 
\newcommand\pasj{\ref@jnl{PASJ}}% 
          % Publications of the ASJ 
\newcommand\qjras{\ref@jnl{QJRAS}}% 
          % Quarterly Journal of the RAS 
\newcommand\skytel{\ref@jnl{S\&T}}% 
          % Sky and Telescope 
\newcommand\solphys{\ref@jnl{Sol.~Phys.}}% 
          % Solar Physics 
\newcommand\sovast{\ref@jnl{Soviet~Ast.}}% 
          % Soviet Astronomy 
\newcommand\ssr{\ref@jnl{Space~Sci.~Rev.}}% 
          % Space Science Reviews 
\newcommand\zap{\ref@jnl{ZAp}}% 
          % Zeitschrift fuer Astrophysik 
\newcommand\nat{\ref@jnl{Nature}}% 
          % Nature 
\newcommand\iaucirc{\ref@jnl{IAU~Circ.}}% 
          % IAU Cirulars 
\newcommand\aplett{\ref@jnl{Astrophys.~Lett.}}% 
          % Astrophysics Letters 
\newcommand\apspr{\ref@jnl{Astrophys.~Space~Phys.~Res.}}% 
          % Astrophysics Space Physics Research 
\newcommand\bain{\ref@jnl{Bull.~Astron.~Inst.~Netherlands}}% 
          % Bulletin Astronomical Institute of the Netherlands 
\newcommand\fcp{\ref@jnl{Fund.~Cosmic~Phys.}}% 
          % Fundamental Cosmic Physics 
\newcommand\gca{\ref@jnl{Geochim.~Cosmochim.~Acta}}% 
          % Geochimica Cosmochimica Acta 
\newcommand\grl{\ref@jnl{Geophys.~Res.~Lett.}}% 
          % Geophysics Research Letters 
\newcommand\jcp{\ref@jnl{J.~Chem.~Phys.}}% 
          % Journal of Chemical Physics 
\newcommand\jgr{\ref@jnl{J.~Geophys.~Res.}}% 
          % Journal of Geophysics Research 
\newcommand\jqsrt{\ref@jnl{J.~Quant.~Spec.~Radiat.~Transf.}}% 
          % Journal of Quantitiative Spectroscopy and Radiative Trasfer 
\newcommand\memsai{\ref@jnl{Mem.~Soc.~Astron.~Italiana}}% 
          % Mem. Societa Astronomica Italiana 
\newcommand\nphysa{\ref@jnl{Nucl.~Phys.~A}}% 
          % Nuclear Physics A 
\newcommand\physrep{\ref@jnl{Phys.~Rep.}}% 
          % Physics Reports 
\newcommand\physscr{\ref@jnl{Phys.~Scr}}% 
          % Physica Scripta 
\newcommand\planss{\ref@jnl{Planet.~Space~Sci.}}% 
          % Planetary Space Science 
\newcommand\procspie{\ref@jnl{Proc.~SPIE}}% 
          % Proceedings of the SPIE

\bibliographystyle{mn2e} 

\bibliography{fornax_dust_v2}
\appendix
\onecolumn

\section{Data Tables}
\begin{table*}
\centering
\begin{tabular}{ccccccccccccc}
\hline
OBJECT &RA &Dec. &$S_{500}$ &$E_{500}$ &$S_{350}$ &$E_{350}$ &$S_{250}$ &$E_{250}$ &$S_{160}$ &$E_{160}$ &$S_{100}$ &$E_{100}$ \\
~ &h:m:s &d:m:s &(Jy) &(Jy) &(Jy) &(Jy) &(Jy) &(Jy) &(Jy) &(Jy) &(Jy) &(Jy) \\
~ &(J2000) &(J2000) &~ &~ &~ &~ &~ &~ &~ &~ &~ &~ \\
\hline

FCC 32 &03:24:52.50 &-35:26:08.0 &0.0 & 0.024 &0.043 & 0.012 &0.094 & 0.015 &-&- &-&- \\
FCC 34 &03:25:02.24 &-35:13:24.2 &0.0 & 0.022 &0.0 & 0.016 &0.0 & 0.015 &-&- &-&- \\
FCC 42 &03:25:46.16 &-35:30:29.5 &0.0 & 0.023 &0.0 & 0.02 &0.0 & 0.017 &-&- &-&- \\
FCC 44 &03:26:07.50 &-35:07:45.0 &0.0 & 0.024 &0.021 & 0.006 &0.015 & 0.004 &0.0 & 0.034 &0.0 & 0.012 \\
FCC 45 &03:26:13.50 &-34:33:15.0 &0.0 & 0.032 &0.0 & 0.023 &0.0 & 0.015 &-&- &-&- \\
FCC 47 &03:26:32.20 &-35:42:50.0 &0.0 & 0.019 &0.0 & 0.017 &0.0 & 0.015 &-&- &-&- \\
FCC 48 &03:26:42.60 &-34:32:57.0 &0.027 & 0.008 &0.031 & 0.008 &0.027 & 0.005 &-&- &-&- \\
FCC 50 &03:26:53.30 &-35:31:12.0 &0.0 & 0.022 &0.0 & 0.017 &0.0 & 0.015 &0.0 & 0.022 &0.0 & 0.019 \\
FCC 55 &03:27:17.60 &-34:31:37.0 &0.0 & 0.023 &0.0 & 0.024 &0.0 & 0.016 &-&- &-&- \\
FCC 56 &03:27:21.50 &-36:08:50.0 &0.0 & 0.021 &0.0 & 0.019 &0.0 & 0.013 &-&- &-&- \\
. &. &. &. &. &. &. &. &. &. &. &. &. \\
. &. &. &. &. &. &. &. &. &. &. &. &. \\
. &. &. &. &. &. &. &. &. &. &. &. &. \\
\hline
\end{tabular}
\captionsetup{width=0.8\textwidth}
\caption{5 band FIR fluxes, uncertainties and upperlimits. If flux density is equal to zero, then $E_{band}$ represents an upper-limit for the galaxy in question, the upper-limit is calculated from the $3\sigma$ noise in the PSF convolved map. Some PACS fluxes are not measured and denoted by a (-) symbol. A machine-readable version of this table is available online with this publication.}
\label{tab:FIRfluxes}
\end{table*}

\newpage

\begin{table*}
\centering
\begin{tabular}{ccccccc}
\hline
OBJECT &RA &Dec. &Type &Dust Temperature &Duss Mass &Stellar Mass \\
~ &h:m:s &d:m:s &~ &K &log($M_{Dust}$/M$_{\odot}$) &log($M_{Stars}$/M$_{\odot}$) \\
~ &(J2000) &(J2000) &~ &~ &~ &~ \\
\hline
FCC 48 &03:26:42 &-34:32:57 &dE &8.78 (1.04) &6.48 (0.26) &7.93 \\
FCC 67 &03:28:48 &-35:10:45 &Sc &17.32 (0.56) &6.86 (0.05) &9.45 \\
FCC 90 &03:31:08 &-36:17:27 &E &20.32 (0.77) &5.6 (0.08) &8.98 \\
FCC 113 &03:33:06 &-34:48:27 &Scd &16.29 (1.43) &5.73 (0.14) &8.88 \\
FCC 121 &03:33:36 &-36:08:17 &Sbc &22.85 (0.44) &8.25 (0.03) &11.16 \\
FCC 167 &03:36:27 &-34:58:31 &S0 &25.77 (0.88) &6.07 (0.04) &10.98 \\
FCC 179 &03:36:46 &-35:59:58 &Sa &23.67 (0.79) &6.78 (0.04) &10.5 \\
FCC 184 &03:36:57 &-35:30:23 &S0 &24.52 (0.69) &6.56 (0.04) &10.77 \\
FCC 215 &03:38:37 &-35:45:27 &dE &15.64 (0.08) &5.23 (0.84) &6.87 \\
FCC 235 &03:40:09 &-35:37:34 &Im &15.78 (1.35) &6.62 (0.12) &9.78 \\
FCC 261 &03:41:21 &-33:46:12 &Irr &11.25 (0.73) &6.13 (0.15) &8.58 \\
FCC 263 &03:41:32 &-34:53:22 &SBcd &21.43 (0.8) &6.21 (0.06) &9.2 \\
FCC 267 &03:41:45 &-33:47:29 &Sm &15.42 (0.82) &5.49 (0.12) &8.48 \\
FCC 282 &03:42:45 &-33:55:13 &Im &18.04 (0.8) &5.83 (0.09) &9.0 \\
FCC 285 &03:43:02 &-36:16:24 &Sd &12.87 (0.94) &6.83 (0.14) &9.38 \\
FCC 290 &03:43:37 &-35:51:14 &Sc &19.08 (0.45) &7.01 (0.04) &10.1 \\
FCC 306 &03:45:45 &-36:20:50 &SBm &13.1 (1.31) &5.79 (0.25) &8.68 \\
FCC 308 &03:45:54 &-36:21:31 &Sd &17.22 (0.53) &6.76 (0.06) &9.39 \\
FCC 312 &03:46:18 &-34:56:33 &Scd &20.35 (0.55) &7.25 (0.04) &10.04 \\
FCC 313 &03:46:33 &-34:41:12 &dS0 &8.76 (0.77) &6.52 (0.19) &7.78 \\
FCC 332 &03:49:49 &-35:56:45 &E &14.93 (1.62) &5.34 (0.24) &8.63 \\
FCC 335 &03:50:36 &-35:54:36 &E &18.55 (0.96) &5.54 (0.11) &9.21 \\
\hline
\end{tabular}
\captionsetup{width=0.8\textwidth}
\caption{22 HeFoCS galaxies dust masses and temperetures given from fitting a modified blackbody ($\beta = 2$ emissivity) to 3 $Herschel$ bands or more.}
\label{tab:FIRmasses}
\end{table*}

\section{SED fits}

\begin{figure*}
\centering
\includegraphics[width=\linewidth]{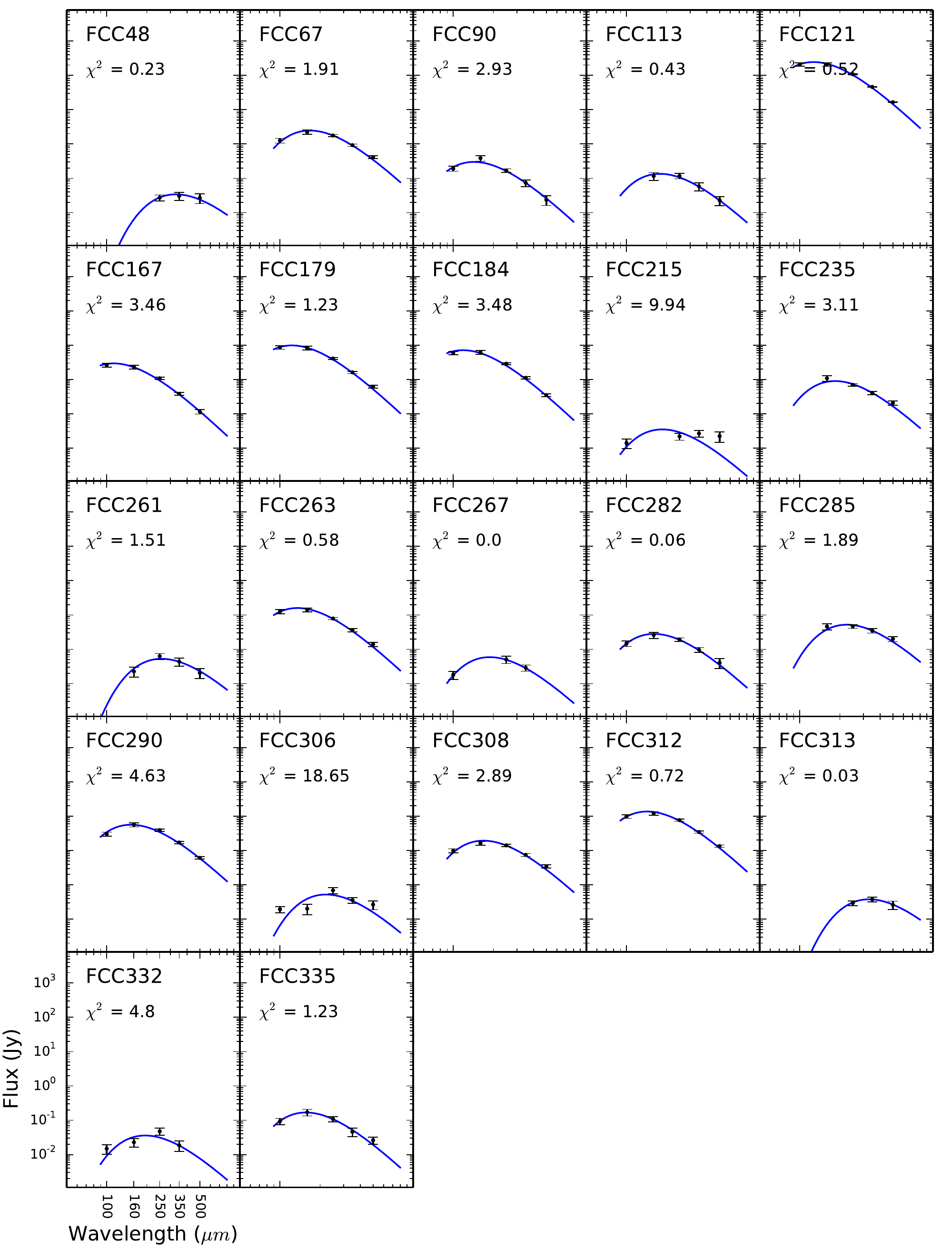}
\caption{Modfied blackbody fits to 22 HeFoCS galaxies. The blue line represents a single tempreture $\beta = 2$ fit to the data. We have only used galaxies with at least 3 $Herschel$ bands. }
\label{fig:sed_fits}
\end{figure*}

\end{document}